\documentclass[aps,prd,floatfix,twocolumn,reprint,amsmath,amssymb,superscriptaddress]{revtex4}
\usepackage{amsfonts}
\usepackage{mathrsfs}
\usepackage{amsmath}% needed for subequations
\usepackage{color}
\usepackage{natbib}
\usepackage{graphicx}
\usepackage{bm}% bold maths
\usepackage{amssymb}
\usepackage{xspace}
\usepackage{epstopdf}
\usepackage{dcolumn}% Align table columns on decimal point
\usepackage{longtable}
\usepackage{multirow}
\usepackage[colorlinks=true, letterpaper=true, pdfstartview=FitV, linkcolor=blue, citecolor=blue, urlcolor=blue]{hyperref}
\usepackage{float}

\makeatletter

\newcommand{\Rmnum}[1]{\expandafter\@slowromancap\romannumeral #1@}
\makeatother
\begin{document}
\title{Nodal surface semimetals: Theory and material realization}

\author{Weikang Wu}
\affiliation{Research Laboratory for Quantum Materials, Singapore University of Technology and Design, Singapore 487372, Singapore}

\author{Ying Liu}
\affiliation{Research Laboratory for Quantum Materials, Singapore University of Technology and Design, Singapore 487372, Singapore}

\author{Si Li}
\affiliation{Research Laboratory for Quantum Materials, Singapore University of Technology and Design, Singapore 487372, Singapore}
\affiliation{Beijing Key Laboratory of Nanophotonics and Ultrafine Optoelectronic Systems, School of Physics, \\ 
	Beijing Institute of Technology, Beijing 100081, China}

\author{Chengyong Zhong}
\affiliation{Institute for Quantum Information and Spintronics, School of Science, \\ 
	Chongqing University of Posts and Telecommunications, Chongqing 400065, China}

\author{Zhi-Ming Yu}
\affiliation{Research Laboratory for Quantum Materials, Singapore University of Technology and Design, Singapore 487372, Singapore}

\author {Xian-Lei Sheng}
\email{xlsheng@buaa.edu.cn}
\affiliation{Department of Physics, Key Laboratory of Micro-nano Measurement-Manipulation and Physics (Ministry of Education), \\
	Beihang University, Beijing 100191, China}
\affiliation{Research Laboratory for Quantum Materials, Singapore University of Technology and Design, Singapore 487372, Singapore}

\author{Y. X. Zhao}
\email{zhaoyx@nju.edu.cn}
\affiliation{National Laboratory of Solid State Microstructures and Department of Physics, Nanjing University, Nanjing 210093, China}
\affiliation{Collaborative Innovation Center of Advanced Microstructures, Nanjing University, Nanjing 210093, China}

\author{Shengyuan A. Yang}
\affiliation{Research Laboratory for Quantum Materials, Singapore University of Technology and Design, Singapore 487372, Singapore}

\begin{abstract}
We theoretically study the three-dimensional topological semimetals with nodal surfaces protected by crystalline symmetries.
Different from the well-known nodal-point and nodal-line semimetals, in these materials, the conduction and valence bands cross on closed nodal surfaces in the Brillouin zone. We propose different classes of nodal surfaces, both in the absence and in the presence of spin-orbit coupling (SOC). In the absence of SOC, a class of nodal surfaces can be protected by spacetime inversion symmetry and sublattice symmetry and characterized by a $\mathbb{Z}_2$ index, while another class of nodal surfaces are guaranteed by a combination of nonsymmorphic two-fold screw-rotational symmetry and time-reversal symmetry. We show that the inclusion of SOC will destroy the former class of nodal surfaces but may preserve the latter provided that the inversion symmetry is broken. We further generalize the result to magnetically ordered systems and show that protected nodal surfaces can also exist in magnetic materials without and with SOC, given that certain magnetic group symmetry requirements are satisfied. Several concrete nodal-surface material examples are predicted via the first-principles calculations. The possibility of multi-nodal-surface materials is discussed.
\end{abstract}
\maketitle

%%%%%%% Main text %%%%%%%%%%%%%%%%%%%%%

\section{Introduction}

Topological states of matter have been attracting great interest in recent physics research. Inspired by the previous studies on topological insulators~\cite{Hasan2010,Qi2011,Bansil2016}, the research focus is now shifted towards topological semimetals~\cite{Chiu2016,Burkov2016,Yan2017,Armitage2017}. In these materials, the electronic band structures possess nontrivial topology and/or symmetry-protected band crossings close to the Fermi level, such that the low-energy electrons behave drastically different from the usual Schr\"{o}dinger-type fermions. For example, in Weyl/Dirac semimetals~\cite{Armitage2017,Wan2011,Murakami2007,Young2012}, the conduction and valence bands cross linearly at isolated $k$ points, around which the electrons resemble the massless Weyl/Dirac fermions from the standard model, making it possible to simulate intriguing high-energy and relativistic physics phenomena in table-top experiments~\cite{Nielsen1983,Volovik-book,Guan2017}.

The nontrivial band crossings can be classified based on their dimensionalities. The crossings in the aforementioned Dirac/Weyl semimetals are isolated zero-dimensional (0D) points. Materials with 1D band crossings, known as nodal-line semimetals, have also been proposed and intensively studied in recent works~\cite{Weng2015,Yang2014,Mullen2015,Yu2015,Kim2015,Chen2015,Fang2016,Li2017Type-PRB}. For 3D materials, there is one remaining possibility: The band crossings may form a 2D nodal surface, namely each point on the surface is a crossing point between two bands whose dispersions are linear along the surface normal direction. Although both are 2D manifolds in the 3D Brillouin zone (BZ), we stress that the nodal surface is distinct from the ordinary Fermi surface, because the coarse-grained quasiparticles excited from a nodal surface have an intrinsic pseudospin degrees of freedom (representing the two crossing bands)~\cite{Zhong2016Towards-N}, behaving effectively as 1D massless Dirac fermions along the surface normal direction, and therefore may have interesting physical properties. In addition, in contrast to the ordinary Fermi surface as a sphere, each nodal surface identified in this article is a torus with opposite edges attached on the periodic boundaries of the BZ.

To date, such nodal-surface semimetals (NSSMs) have only appeared in a few scattered theoretical studies. In Ref.~\cite{Zhong2016Towards-N}, Zhong \emph{et al.} proposed a family of stable graphene network materials, each exhibiting a pair of nodal surfaces close to the Fermi energy. Liang \emph{et al.}~\cite{Liang2016Node-PRB} studied the hexagonal Ba$MX_3$ ($M=$ V, Nb, or Ta; $X=$ S or Se) and found a single nodal surface in these materials when the spin-orbit coupling (SOC) can be neglected. From symmetry and topology analysis, Bzdu\v{s}ek and Sigrist discussed the possibility to stabilize nodal surfaces in centrosymmetric systems~\cite{Bzdusek2017}. More recently, the nodal surface similar to Ref.~\cite{Liang2016Node-PRB} was predicted for a proposed acoustic metamaterial~\cite{MXiao2017}, and the stability of nodal surfaces against perturbations was theoretically investigated~\cite{Turker2017}.  Admittedly, our current understanding of NSSMs is still at the primitive stage with several important issues to be addressed. First, the nodal surfaces in the carbon allotropes were found through first-principles calculations, but its topological origin has not been clearly elucidated. Second, given that the nodal surfaces in the aforementioned works are all vulnerable against SOC, \emph{is it possible to have nodal surfaces robust under SOC?} This important question is still waiting to be answered. Third, the carbon allotropes in Ref.~\cite{Zhong2016Towards-N} have yet to be synthesized, while the nodal surfaces in BaVS$_3$~\cite{Liang2016Node-PRB} can only be maintained in the room-temperature structural phase. It is thus important to search for more realistic material candidates to facilitate the experimental investigation on NSSMs.

Motivated by the above questions and issues, in this work, we present a theoretical study of the 3D NSSMs both in the absence and in the presence of SOC. In the absence of SOC, we propose two different classes of symmetry-protected nodal surfaces corresponding to the two materials studied in Refs.~\cite{Zhong2016Towards-N,Liang2016Node-PRB}. The first class of the nodal surfaces (as those in the graphene networks) are protected by spacetime inversion symmetry and sublattice symmetry, and are characterized by a $\mathbb{Z}_2$ topological index. The second class of nodal surfaces are guaranteed by a combination of two-fold screw-rotational symmetry and time-reversal symmetry. After the inclusion of SOC, the former class of nodal surfaces is destroyed generically, but the latter can still be protected with the requirement that the space-time inversion symmetry is violated. Furthermore, it is found that both in the absence and in the presence of SOC, the protection in the latter case persists even in the presence of magnetic orders perpendicular to the screw axis, where, although time reversal symmetry is violated, the system is still invariant under the combination of screw rotation and time reversal. Via first-principles calculations, we identify several candidate NSSM materials with relatively clean low-energy band structures, which support the theoretical analysis in this paper and will facilitate experimental studies on the novel physical properties associated with nodal surfaces.

It is noteworthy that the band structures of the two classes are essentially different due to their qualitatively distinct origins.  Nodal surfaces in the first classes have the nontrivial $\mathbb{Z}_2$ topological charge, and therefore appear in pairs in the BZ conforming the Nielsen-Ninomiya no-go theorem~\cite{Nielsen1981,Nielsen1981a}, while those in the second class mainly due to the two-fold nonsymmorphic symmetry exist alone in the BZ, for which the topological essence was revealed in Ref.~\cite{Zhao-Schnyder-Nonsymm}. Particularly, each degenerate point on such a nodal surface belong to a twisted band structure analogous to the M\"{o}bius strip~\cite{Zhao-Schnyder-Nonsymm}, and the band structure as a whole is equivalent to a collection of Mobius strip parametrized by a torus (a 2D sub-BZ perpendicular to the screw axis).

This paper is organized as follows. In Sec.~II, we discuss the two classes of nodal surfaces in the absence of SOC. In Sec.~III, we propose nodal surfaces that can be protected in the presence of SOC. In Sec.~IV, we generalize the discussion to magnetically ordered systems, predicting nodal surfaces in magnetic materials. In each section, we present corresponding material examples with band structures obtained from first-principles calculations. The details of our first-principles methods are presented in the Appendix. Discussion and conclusion are presented in Sec.~V.

\section{Nodal surface in the absence of SOC}\label{S2}

\subsection{Nodal surface with $\mathbb{Z}_2$ topological charge}\label{2A}

Let us start with the first class (class I) of nodal surfaces, which require spacetime inversion symmetry $\mathcal{P}\mathcal{T}$ with the relation $(\mathcal{P}\mathcal{T})^2=1$ and sublattice symmetry $\mathcal{S}$. Here $\mathcal{P}$ and $\mathcal{T}$ denote space inversion and time reversal, respectively, and the sublattice symmetry $\mathcal{S}$ satisfies the anti-commutation relation with the Hamiltonian: $\{\mathcal{S},\mathcal{H}(\bm k)\}=0$. It is required that $\mathcal{P}\mathcal{T}$ and $\mathcal{S}$ are mutually independent, namely $[\mathcal{P}\mathcal{T},\mathcal{S}]=0$. The relation of $(\mathcal{P}\mathcal{T})^2=1$ can be realized by particles with integer spin, such as photonic crystals or phonons of mechanical systems and certain cold-atom systems and by electronic systems with negligible SOC, such as the carbon allotropes studied in Ref.~\cite{Zhong2016Towards-N}. In Ref.~\cite{Zhong2016Towards-N}, the symmetry/topology aspect was not fully exposed, and will be our focus in the following discussion.

We now address the $\mathbb{Z}_2$ topological classification of the class-I nodal surfaces under the constraint of $\mathcal{PT}$ and $\mathcal{S}$ symmetries~\cite{PT-Classification,Bzdusek2017}. Since both $\mathcal{P}$ and $\mathcal{T}$ inverse $\bm{k}$, $\mathcal{PT}$ operates trivially in momentum space. As $\mathcal{P}$ is unitary but $\mathcal{T}$ anti-unitary, we can choose the representation in momentum space that $\mathcal{PT}={\mathcal{K}}$, which is unique up to a unitary transformation. Here $\mathcal{K}$ is the complex conjugation. Therefore, a $\mathcal{PT}$-symmetric Hamiltonian $\mathcal{H}(\bm{k})$ is real in this representation. On the other hand, if $\mathcal{S}$ is represented by $\mathcal{S}=\sigma_z\otimes 1_N$, where $\sigma_z$ is the third Pauli matrix and $1_N$ the $N\times N$ identity matrix with $N$ being the number of valence bands, then the Hamiltonian $\mathcal{H}(\bm{k})$ takes the block anti-diagonal form, i.e., $\mathcal{H}(\bm{k})=\mathrm{antidiag}[Q(\bm{k}),Q^\dagger(\bm{k})]$~\cite{PT-Classification}. Because of $[\mathcal{PT},\mathcal{S}]=0$, $Q(\bm k)$ can always be converted to a real matrix by a unitary transformation. Thus, for any gapped momentum $\bm{k}$, the Hamiltonian $\mathcal{H}(\bm{k})$ (after being flattened) can be topologically regarded as a point in the space $O(N)$ (with the assumption $N\ge 2$), which is exactly the Hamiltonian space for the class BDI among the ten Altland-Zirnbauer (AZ) symmetry classes~\cite{AZ-classes,TI-classification}. The space of $O(N)$ has two disconnected components, namely $\pi_0[O(N)]=\mathbb{Z}_2$, and the sign of the determinant of $Q(\bm{k})$ specifies which component $\mathcal{H}(\bm{k})$ belongs to.

In the framework of topological classification of nodal surfaces, the \emph{spatial} codimension of nodal surfaces in three dimensions is zero, namely a 0D sphere $S^0$ consisting of two points is chosen to surround a nodal surface in 3D BZ from its transverse dimensions~\cite{Volovik-book,Zhao-Wang-Classification}. We now consider two gapped points $\bm{k}_{1,2}$ with $\mathrm{sgn}[\mathrm{Det}Q(\bm{k}_{1,2})]=\pm 1$, respectively. Then, any path connecting the two points in momentum space has to pass through at least a band-crossing point $\bm{k}_0$ with $\mathrm{Det}Q(\bm{k}_0)=0$ as a result of the mean value theorem, and the band-crossing point has a two-fold degeneracy, which is of topological stability once $\mathcal{PT}$ and $\mathcal{S}$ are preserved. These two-fold degenerate band-crossing points generically spread out a compact nodal surface in momentum space, namely a class-I nodal surface, which accordingly has a non-trivially $\mathbb{Z}_2$ topological charge.

\begin{figure}[hbt]
\includegraphics[width=0.48\textwidth]{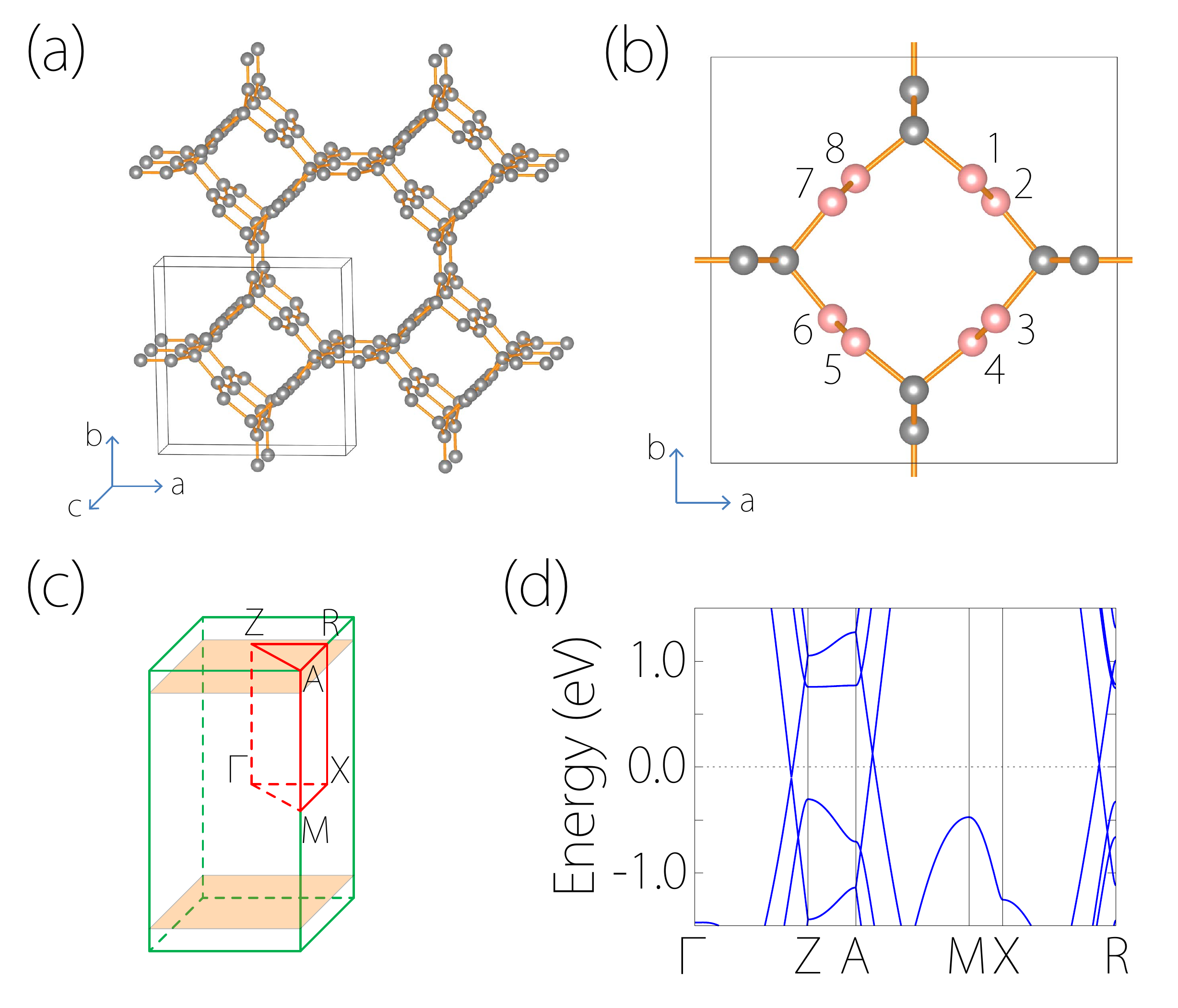}
\caption{(a) Lattice structure and (b) unit cell of the carbon allotrope QGN(1,2). (a) shows a $2 \times 2 \times 2$ supercell, in which the unit cell is marked by the black box. In (b), the red-colored carbon atoms have a $sp^2$ hybridization character (which are labeled from site 1 to site 8), whereas the blue-colored carbon atoms have a $sp^3$ hybridization character. (c) Corresponding Brillouin zone, where the two nodal surfaces are schematically shown by the shaded surfaces. (d) Calculated band structure for QGN(1,2) without SOC.}
\label{QGN}
\end{figure}

The class-I nodal surface can be realized by the carbon allotropes, which has been studied in Ref.~\cite{Zhong2016Towards-N} with other emphases. This family of materials share both $\mathcal{P}$ and $\mathcal{T}$ symmetry with negligible SOC, and therefore have the desired $\mathcal{PT}$ symmetry with $(\mathcal{PT})^2=1$. The chiral symmetry $\mathcal{S}$ also emerges in the low-energy effective theory by appropriately tuning the chemical potential of each site. To be concrete, we take one particular example in this family, namely QGN(1,2), where QGN stands for the quadrilateral graphene network, as shown in Fig.~\ref{QGN}. The crystal structure of QGN(1,2) has the space group symmetry of $P4_2/mmc$ (No.~131). Its unit cell contains two kinds of graphene nanorbbon-like structural motifs: The edge atoms show a strong $sp^3$ hybridization character, whereas the remaining atoms exhibit $sp^2$ hybridization character and form four zigzag chains running along the $c$-axis. From atomic orbital projection, one finds that the low-energy states are dominated by the $\pi$ orbitals on the eight $sp^2$ carbon atomic sites which are marked in red color in Fig.~\ref{QGN}(b). Thus, it is emerged a sublattice symmetry in low-energy regime, which relates electrons at odd sites to holes at their even nearest neighbours and vice versa (see the detailed discussion below), given that the chemical potential is appropriately tuned for each orbital.

The band structure of QGN(1,2) obtained from the first-principles calculations is plotted in Fig.~\ref{QGN}(d). The band structures with and without SOC have little difference, due to the negligible SOC strength of carbon atoms. One observes the linear-type band crossing along the $\Gamma$-Z and the M-A paths. As mentioned in Ref.~\cite{Zhong2016Towards-N}, these crossing points form nodal surfaces and there is a pair of such surfaces related by $\mathcal{T}$ or $\mathcal{P}$ in the BZ, as illustrated in Fig.~\ref{QGN}(c). Each point on the surfaces is a linear crossing point between two bands along the surface normal direction (approximately the $k_z$ direction). And either nodal surface is quite flat with very small energy variation, and is close to the Fermi level.

To reveal the topological nature of the class-I nodal surface of QGN(1,2), we first work out the tight-binding (TB) model corresponding to its low-energy physics. At low energies, the two crossing bands are dominated by the $\pi$ orbitals from the eight $sp^2$ carbon atoms in a unit cell, which are numbered clockwisely from 1 to 8 as shown in Fig.~\ref{QGN}(b), and the eight-band TB model $\mathcal{H}(\bm{k})$ is explicitly given in Appendix~\ref{TB-model}. The sublattice symmetry $\mathcal{S}$ relates $c_i$ to $c_{i+1}^\dagger$ with $i=1,3,5,7$, and therefore is represented as $\mathcal{S}=1_4\otimes\sigma_z$. Here and hereafter, we express the $8\times 8$ matrices using the Kronecker products of the Pauli matrices ($\sigma_i$, $\tau_i$) and the identity matrices. The chiral symmetry operator can be converted to $\widetilde{\mathcal{S}}=\sigma_z\otimes 1_4$ by the unitary transformation $U_\Gamma$ that exchanges $c_2$ and $c_7$ as well as $c_4$ and $c_5$, so that the transformed Hamiltonian is block diagonalized with the upper right block denoted by $A(\bm{k})$. The operators are now ordered as $c=(c_1,c_7,c_3,c_5,c_4,c_6,c_2,c_8)^T$. From Fig.~\ref{QGN}(b), $\mathcal{P}$ maps $c_1$ to $c_5$, $c_3$ to $c_7$ for odd orbitals, and $c_2$ to $c_6$, $c_4$ to $c_8$ for even orbitals, namely $\mathcal{P}=1_2\otimes(\sigma_x\otimes\tau_x)$ with $\tau_x$ being the first Pauli matrix. Since $\mathcal{T}$ is simply complex conjugation, it is found that $\mathcal{PT}=1_2\otimes\sigma_x\otimes\tau_x {\mathcal{K}}$ , which commutes with $\widetilde{\mathcal{S}}$. The relation $(\mathcal{PT})^2=1$ implies there exists a unitary transformation converting $\mathcal{PT}$ to $\widetilde{\mathcal{PT}}={\mathcal{K}}$, which turns out to be $U_{PT}=1_2\otimes e^{i\frac{\pi}{4}\sigma_x\otimes\tau_x}$. After this unitary transformation, the upper right block $A(\bm{k})$ becomes a real matrix $Q(\bm{k})$. Now the $\mathbb{Z}_2$ topological charge of a nodal surface can be defined as
\begin{equation}\label{Z2}
\nu=(\mathrm{sgn}\mathrm{Det}[Q(\bm{k}_1)]-\mathrm{sgn}\mathrm{Det}[Q(\bm{k}_2)])/2\mod 2,
\end{equation}
where $\bm{k}_{1,2}$ are two points on two sides of the nodal surface, respectively. It is verified in Appendix~\ref{TB-model} that both nodal surfaces illustrated in Fig.~\ref{QGN}(c) have the nontrivial topological charge.

We have two remarks before proceeding. First, in Ref.~\cite{Bzdusek2017}, Bzdu\v{s}ek and Sigrist developed a theory for the centrosymmetric extension of the ten AZ classes. The example discussed here would fit into the BDI class discussed in their work. Second, the similar $\mathcal{S}$ symmetry (known as the chiral symmetry) also naturally emerge in the low-energy excitation spectrum for superconductors. Hence, nodal surfaces may also appear in certain centrosymmetric superconductors, like that discussed in Refs.~\cite{Agterberg2017,Bzdusek2017}.

\subsection{Essential nodal surface dictated by nonsymmorphic symmetry}\label{2B}
Next, we consider the second class (class-II) of nodal surfaces existing in the absence of SOC. Such surface is a result of the combination of a two-fold screw symmetry and time reversal symmetry. It is \emph{essential} in the sense that its presence and location are solely dictated by the symmetries, as we discuss below.

Consider the two-fold screw rotation $S_{2z}:\; (x,y,z)\rightarrow (-x,-y,z+\frac{1}{2})$, which is a nonsymmorphic symmetry involving a half translation along the rotation axis. In momentum space $S_{2z}$ inverses $k_x$ and $k_y$, and preserves $k_z$. Without SOC, one finds that
$
(S_{2z})^2=T_{001}=e^{-ik_z}
$,
where $T_{001}$ is the translation along the $z$-direction by a lattice constant. $\mathcal{T}$ is anti-unitary and inverses $\bm{k}$ with the relation $\mathcal{T}^2=1$. Thus, the combination $\mathcal{T}S_{2z}$ is anti-unitary and only inverses $k_z$. Since $[\mathcal{T},S_{2z}]=0$, $\mathcal{T}S_{2z}$ satisfies
\begin{equation}
(\mathcal{T}S_{2z})^2=e^{-ik_z}. \label{magnetic-nonsymm}
\end{equation}
We now regard the system under consideration as a collection of $1$D $k_z$-subsystems  parametrized by $(k_x,k_y)$. Each $1$D subsystem with given $(k_x,k_y)$ effectively has the same non-symmorphic magnetic symmetry $\mathcal{T}S_{2z}$ with the relation \eqref{magnetic-nonsymm}. Hence for a two-band theory, it has a twisted band structure of M\"{o}bius strip with a single band-crossing at the boundary of the 1D sub-BZ, which is enforced by the topological nature of the two-fold non-symmorphic symmetry, as shown in Refs.~\cite{Zhao-Schnyder-Nonsymm,Michel1999,Young2015}, where the momentum-dependent symmetry operator is also associated with a unit winding number.  The M\"{o}bius-strip type band structure has a nontrivial $\mathbb{Z}_2$ topological charge, which means that, given the symmetry being preserved, a single band-crossing is topologically stable, and can only be gapped in pair in four-band theories~\cite{Zhao-Schnyder-Nonsymm}. Since a linear crossing point exists for every 1D subsystem $(k_x,k_y)$ at the sub-BZ boundary, the collection of crossings form a nodal surface on the boundary plane of the 3D BZ with $k_z=\pi$.

The crossings on the surface can also be understood as a result of the Kramers degeneracy. One notes that \emph{any} point on the $k_z=\pi$ plane is invariant under $\mathcal{T}S_{2z}$, but from Eq.~\eqref{magnetic-nonsymm}, the anti-unitary symmetry satisfies
\begin{equation}\label{TS2}
(\mathcal{T}S_{2z})^2=-1
\end{equation}
on the whole plane. Thus, the two-fold Kramers degeneracy arises at every point on the plane. This degeneracy is generically lifted away from this plane due to the loss of symmetry protection, so that a nodal surface is formed at the $k_z=\pi$ plane. The above argument was presented in Ref.~\cite{Liang2016Node-PRB}, where BaVS$_3$ was identified as a candidate material with this type of nodal surfaces.

As we have mentioned, the condition for the presence of the class-I nodal surface is quite stringent: Besides the symmetries $\mathcal{PT}$ and $\mathcal{S}$, its appearance also requires regions in BZ with inverted band orderings (different $\mathbb{Z}_2$ indices). In contrast, the presence of the class-II nodal surface discussed here is solely guaranteed by symmetry (hence can be regarded as an essential band-crossing), and its location is fixed at $k_z=\pi$ (if the screw axis is along $z$). This makes it easier for search for candidate materials by analyzing the space groups.

We emphasize that although the existence of class-II nodal surface is determined by symmetry, its energy is not determined and depends on the specific material. In addition, there could be strong energy variation from point to point on the nodal surface. In order for the nodal surface to manifest in physical properties, a ``good" candidate NSSM needs to satisfy the following requirements: (i) the nodal surface should have relatively small energy variation; (ii) its energy should be close to Fermi level; and (iii) it is desired that no other extraneous band is present at low energy.

\begin{figure}[hbt]
\includegraphics[width=0.46\textwidth]{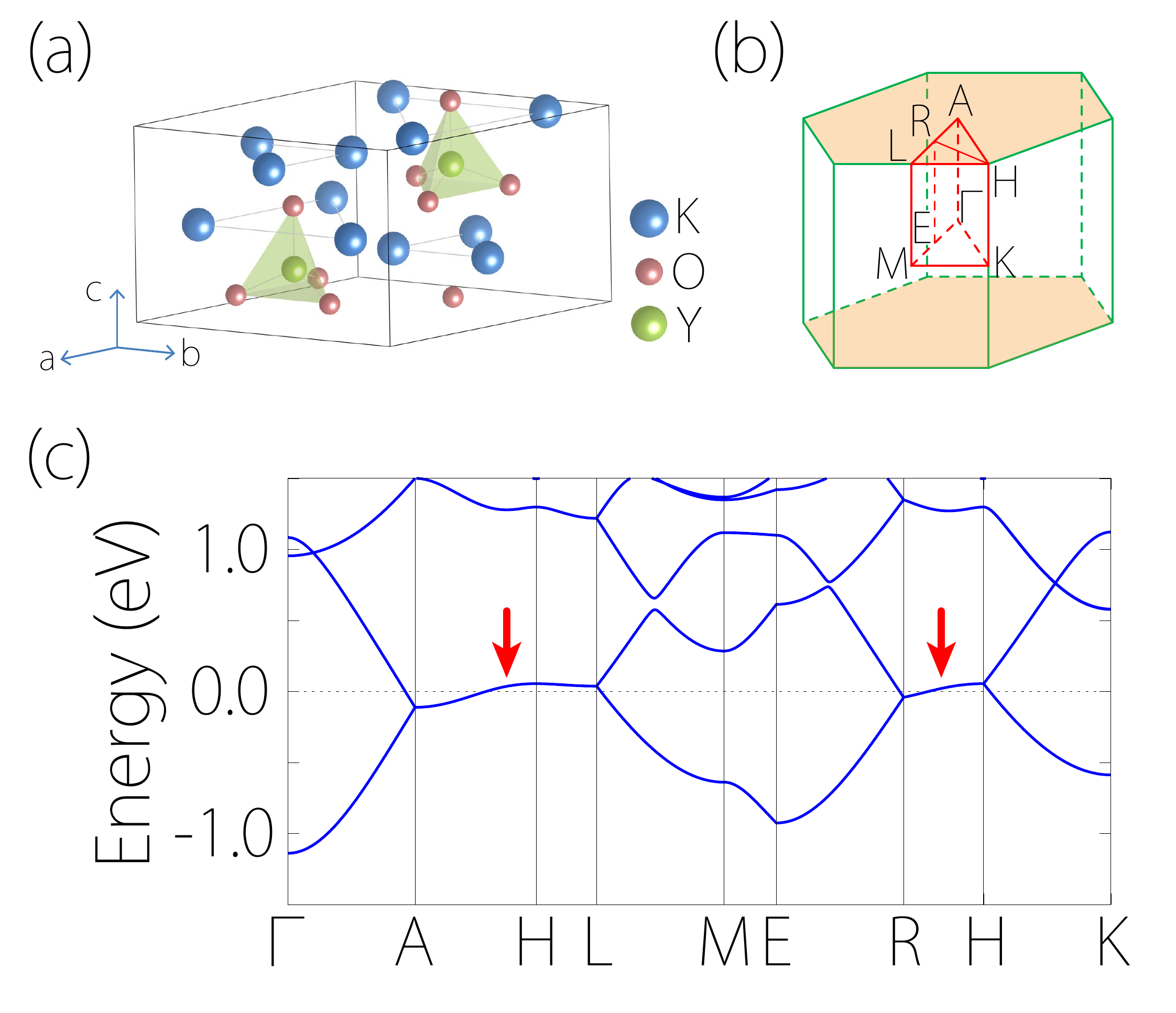}
\caption{(a) Crystal structure of K$_{6}$YO$_{4}$ and (b) the corresponding Brillouin zone. In (b), R and E are the mid-points of the paths A-L and $\Gamma$-M, respectively. (c) Band structure of K$_{6}$YO$_{4}$ without SOC. The red arrows indicate the band degeneracy along the paths on the nodal surface. }
\label{K6YO4}
\end{figure}

Here we identify example materials which possess the class-II nodal surfaces. The first example is the compound K$_{6}$YO$_{4}$ with the structure in the space group $P6_{3}mc$ (No.~186), as illustrated in Fig.~\ref{K6YO4}(a). This material is predicted in Materials Project~\cite{Jain2013Materials-AM}. In the crystal structure, each yttrium atom is surrounded by a tetrahedron of O atoms and is located at the $2b$ positions: $(\frac{1}{2},\frac{2}{3},u)$ and $(\frac{2}{3},\frac{1}{3},u+\frac{1}{2})$ with $u=0.239$, while the K atoms fill the space between the tetrahedra, forming two types of triangles with different sizes. The material is nonmagnetic and has a two-fold screw axis along the ${z}$-direction exists, hence the condition for a class-II nodal surface is satisfied. To be noted, the material does not have the inversion symmetry.

The calculated band structure of K$_{6}$YO$_{4}$ in the absence of SOC is displayed in Fig.~\ref{K6YO4}(c). Since the material is composed of light elements, the SOC has negligible effect on the band structure, which has been checked by our density functional theory (DFT) calculation. From the result, one observes that a nodal surface is indeed present in the $k_z=\pi$ plane, where the two low-energy bands cross linearly. In Fig.~\ref{K6YO4}(c), we purposely show the dispersion along a generic $k$-path E-R, which is not a high-symmetry path [see Fig.~\ref{K6YO4}(b)] and on which the linear crossing can be clearly observed.
This is in agreement with the above argument that the anti-unitary symmetry $\mathcal{T}S_{2z}$ guarantees a nodal surface at the $k_z=\pi$ plane.  In addition, the energy variation of the nodal surface at the $k_{z}=\pi$ plane is small ($<0.2$ eV), the linear dispersion range is relatively large (above 0.5 eV), and there is no other extraneous band close to the Fermi level. These features are desired for a NSSM.

\begin{figure}[hbt]
\includegraphics[width=0.48\textwidth]{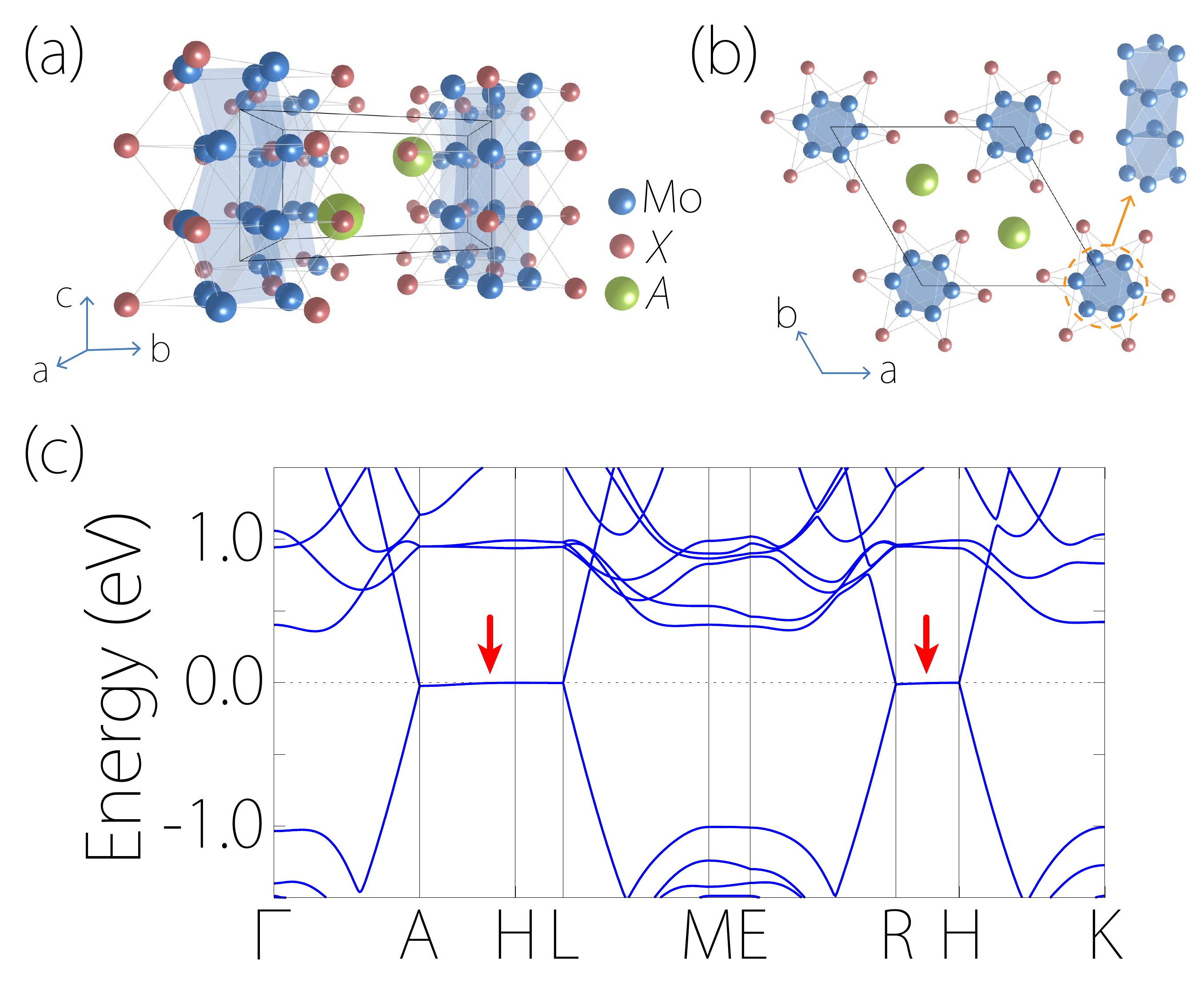}
\caption{(a) Perspective view and (b) top view of the crystal structure of the $A$Mo$_{3}X_{3}$ family materials, where $A =$ (Na, K, Rb, In, Tl) and $X =$ (S, Se, Te). The insert in (b) shows the face-sharing Mo$_{6}$ octahedra chains. (c) Band structure of RbMo$_{3}$S$_{3}$ without SOC. The red arrows indicate the band degeneracy along the paths on the nodal surface. The Brillouin zone here has the same shape as that in Fig.~\ref{K6YO4}(b).}
\label{AMo3X3}
\end{figure}

The second example is a family of materials with the formula $A$Mo$_{3}X_{3}$, where $A =$ (Na, K, Rb, In, Tl) and $X =$ (S, Se, Te). These $A$Mo$_{3}X_{3}$ compounds have been synthesized in experiment~\cite{Hoenle1980New-JotLCM,Huster1983Neue-JotLCM}. They share the same crystalline structure with the space group $P6_{3}/m$ (No.~176), as shown in Figs.~\ref{AMo3X3}(a) and~\ref{AMo3X3}(b). The structure features close-packed one-dimensional $[\text{Mo}_{3}X_{3}]$ columns, consisting of face-sharing Mo$_{6}$ octahedra surrounded by $X$ atoms. The $[\text{Mo}_{3}X_{3}]$ columns are oriented along the $z$-direction, arranged in a trigonal lattice in the $x$-$y$ plane, and related to each other by a screw rotation. The $A$ atoms are intercalated in the large holes between the columns. The space group contains the $S_{2z}$ and $\mathcal{T}$ symmetries. And it should be noted that the structure preserves the inversion symmetry $\mathcal{P}$. This point will be important for the later discussion in Sec.~III when we deal with the case with SOC included.

Let us consider one specific example RbMo$_{3}$S$_{3}$ in this family, for which the effect of SOC on the low-energy bands is small. The calculated band structure for RbMo$_{3}$S$_{3}$ in the absence of SOC is shown in Fig.~\ref{AMo3X3}(c). Similarly to K$_{6}$YO$_{4}$, one observes a nodal surface located at the $k_z=\pi$ plane, guaranteed by the anti-unitary symmetry $\mathcal{T}S_{2z}$. The linear dispersion range is above 1 eV, even larger than that of K$_{6}$YO$_{4}$. The nodal surface is very flat in energy, and it sits almost exactly at the Fermi level. Such a band structure would be ideal for studying the class-II nodal surfaces.

Before proceeding, we mention that the structure of the $A$Mo$_{3}X_{3}$ family compounds have strong quasi-1D character, which may induce structural instabilities towards Peierls distortion. First-principles calculations indicated that except for TlMo$_3$Te$_3$ and RbMo$_3$Te$_3$, other members in this family would be prone to a Peierls distortion that breaks the $S_{2z}$ symmetry~\cite{Liu2017}, hence destroying the nodal surface. However, such distortion has not been detected in experiment~\cite{Potel1980,Chevrel1986}, and future studies are needed to clarify this issue.

\section{Nodal surface in the presence of SOC}\label{S3}

Compared to the case without SOC, the inclusion of SOC makes at least two important differences in terms of symmetry properties: (i) The time reversal operator now satisfies $\mathcal{T}^2=-1$ because the time reversal operation reverses the electron spin and (ii) all rotation operations need to operate on the spin in addition to the spatial degree of freedom.

For class-I nodal surfaces, the protection mechanism discussed in Sec.~\ref{2A} no longer holds, because of point (i). Hence introducing SOC will generally destroy class-I nodal surfaces. We explicitly verify this point by artificially increasing the SOC strength in QGN(1,2), and indeed find that the nodal surface is gapped out by SOC (see Appendix~\ref{appC}).

For class-II nodal surfaces, a more involved analysis is needed. It turns out that the inversion symmetry $\mathcal{P}$ plays an important role in this case. In order to ensure the boundary Kramers degeneracies due to $\mathcal{T}S_{2z}$ are protecting a nodal surface, $\mathcal{PT}$ symmetry has to be violated. First, Eq.~\eqref{TS2} still holds with SOC. Although $\mathcal{T}^2=-1$, performing $S_{2z}$ twice also rotates spins, and we have
\begin{equation}
(S_{2z})^2=T_{001}\bar{E}=-e^{-ik_z},
\end{equation}
where $\bar{E}$ denotes the $2\pi$-rotation on spin, contributing a factor of $-1$. As $[\mathcal{T},S_{2z}]=0$, Eq.~\eqref{magnetic-nonsymm} still holds, so does Eq.~\eqref{TS2}. However, as long as inversion symmetry is present, the Kramers degeneracies on the BZ boundary with $k_z=\pi$ can extend to the whole bulk of BZ (along any generic path) because of the spacetime inversion symmetry $\mathcal{PT}$. In contrast to the case without SOC, $(\mathcal{PT})^{2}=-1$, hence there is actually the Kramers degeneracy at every point in the BZ due to $\mathcal{PT}$, which acts locally in momentum space. Consequently, the degeneracy at the $k_z=\pi$ plane in this case does not represent a nodal surface according to our definition.

\begin{figure}[hbt]
\includegraphics[width=0.48\textwidth]{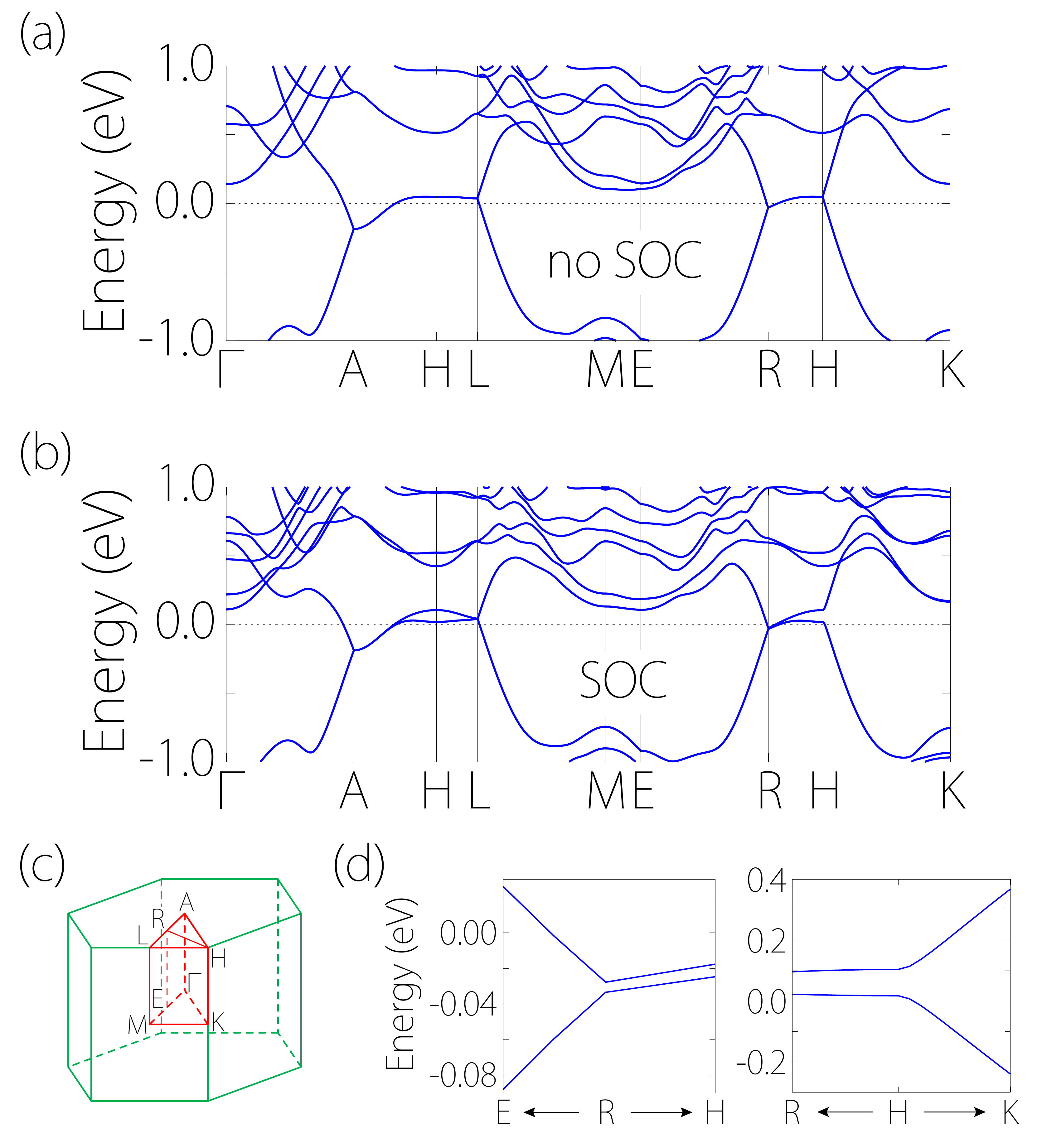}
\caption{Calculated band structures of TlMo$_{3}$Te$_{3}$ (a) without SOC and (b) with SOC. Panel (c) shows the Brillouin zone. R and E are the mid-points of the paths A-L and $\Gamma$-M, respectively. (d) Enlarged view of the low-energy band dispersion around the R and H points, showing that the original band-crossing at the nodal surface is gapped by SOC.}
\label{TlMo3Te3}
\end{figure}

To explicitly demonstrate the above point, we perform the first-principles calculation on the material TlMo$_{3}$Te$_{3}$ in the $A$Mo$_{3}X_{3}$ family with SOC included. Its crystal structure has been shown in Fig.~\ref{AMo3X3}(a) and~\ref{AMo3X3}(b). As mentioned before, the lattice structure of TlMo$_{3}$Te$_{3}$  preserves the inversion symmetry $\mathcal{P}$, and unlike RbMo$_{3}$S$_{3}$, the SOC in TlMo$_{3}$Te$_{3}$ has appreciable effects on the band structure and needs to be taken into account. In Figs.~\ref{TlMo3Te3}(a) and \ref{TlMo3Te3}(b), we plot the band structures of TlMo$_{3}$Te$_{3}$ without and with SOC. In the absence of SOC, a nodal surface is present at the $k_z=\pi$ plane, similar to that for RbMo$_{3}$S$_{3}$. After including SOC, band splitting is observed at the original nodal surface. One notes that the band dispersion along the K-H and E-R no longer shows the band crossing [see Figs.~\ref{TlMo3Te3}(b) and \ref{TlMo3Te3}(d)], hence the two-fold degeneracy by $\mathcal{T}S_{2z}$ in the $k_z=\pi$ plane does not lead to a nodal surface. This demonstrates that when $\mathcal{P}$ is preserved, $\mathcal{T}S_{2z}$ can no longer guarantee a nodal surface in the presence of SOC. Meanwhile, it is noted that in Fig.~\ref{TlMo3Te3}(b), the crossing is maintained along $\Gamma$-A path at the A point, which is enabled by the six-fold rotational symmetry on this path. A recent study shows that the crossing point at A is actually a cubic Dirac point with linear dispersion along $k_z$ and cubic dispersion in the $k_z=\pi$ plane~\cite{Liu2017}.

\begin{figure}[hbt]
\includegraphics[width=0.49\textwidth]{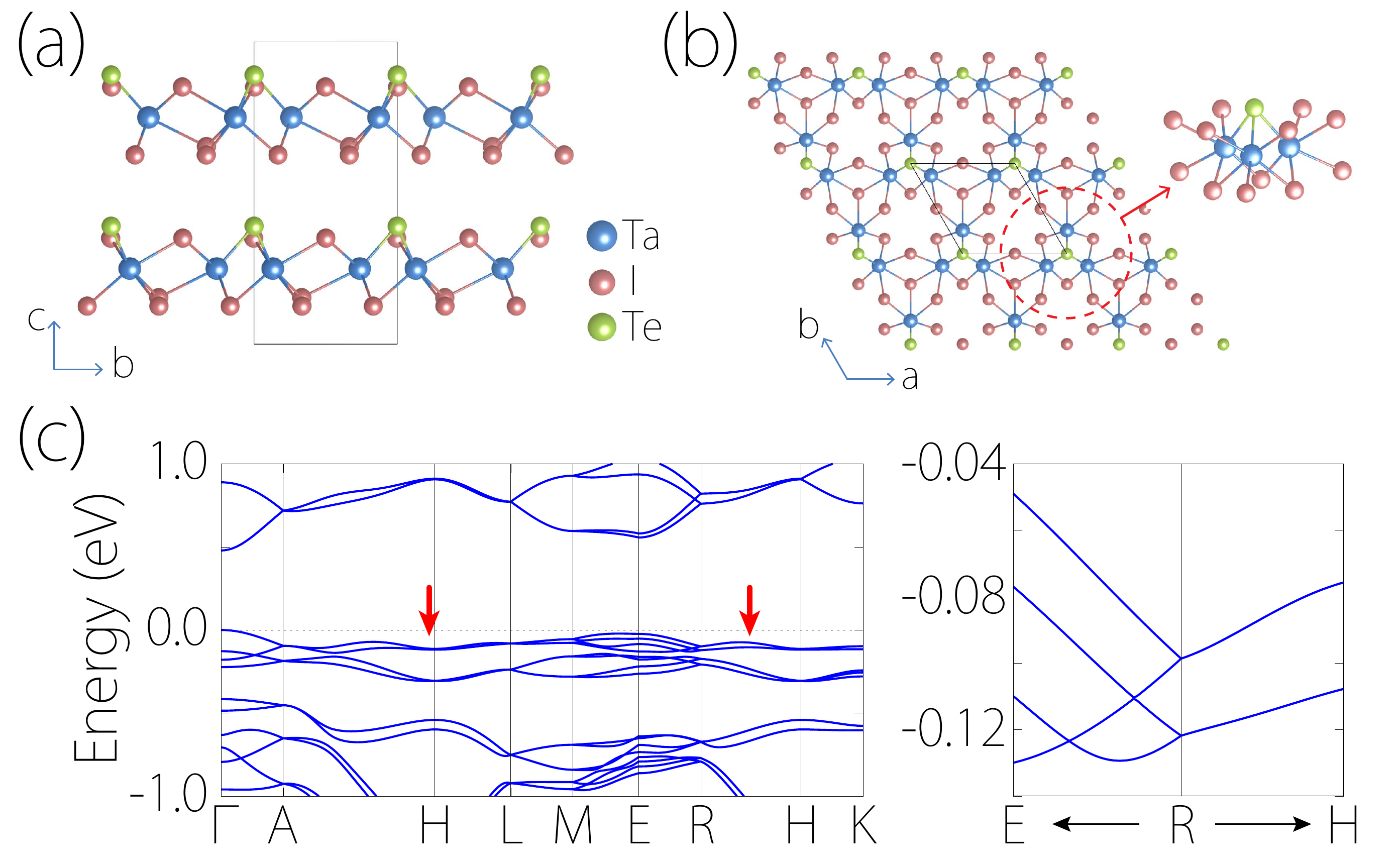}
\caption{(a) Side view and (b) top view of the crystal structure for Ta$_{3}$TeI$_{7}$. The inset in (b) shows the cluster unit of the triangular lattice. (c) Calculated band structure for Ta$_{3}$TeI$_{7}$ in the presence of SOC. The right panel shows the enlarged view of the band structure near the Fermi level along a generic path E-R-H, where E-R is perpendicular to the $k_z=\pi$ plane and R-H is in the $k_z=\pi$ plane. The Brillouin zone and the labeled $k$ points are the same as in Fig.~\ref{K6YO4}(b).}
\label{Ta3TeI7}
\end{figure}

Hence, from the above argument, a necessary condition for a $\mathcal{T}S_{2z}$-protected nodal surface would be that the $\mathcal{P}$ symmetry needs to be broken (here we consider nonmagnetic materials where $\mathcal{T}$ is preserved). This will generally lift the two-fold degeneracy away from the $k_z=\pi$ plane, thus pairs of non-degenerate bands along a generic path have to cross at the $k_z=\pi$ plane, making the doubly-degenerate $k_z=\pi$ plane a nodal surface. Here, it should be mentioned that for systems without SOC (those in Sec.~\ref{2B}), spin is a dummy degree of freedom. Hence, if the spin degeneracy is counted, then the nodal surface there would be four-fold degenerate. In comparison, in the presence of SOC, the obtained nodal surfaces are two-fold degenerate essentially due to a Kramers-like degeneracy, and there is no protection for a four-fold degenerate surface (at least by the symmetries discussed in this work).

This picture is indeed confirmed by our calculation on an example material Ta$_{3}$TeI$_{7}$. Its crystal structure has the space group $P6_{3}mc$ (No.~186), which does not possess an inversion center. Ta$_{3}$TeI$_{7}$ is a member of the $M_{3}QX_{7}$ material family ($M =$ Nb, Ta; $Q =$ S, Se, Te; and $X =$ Cl, Br, I)~\cite{Smith1996Novel-JotACS,Smith1998Ta3SBr7A-JoSSC}. The crystal structure consists of ordered, close-packed layers of I and Te atoms, interleaved by Ta atoms, as shown in Fig.~\ref{Ta3TeI7}(a) and~\ref{Ta3TeI7}(b). Figure~\ref{Ta3TeI7}(b) displays the top view of a single Ta$_{3}$TeI$_{7}$ layer,  which is similar to the structure of 1T'-MoS$_{2}$. This layered material have several types of stacking. Here we consider the bulk structure with the $ABAB$ layer stacking pattern. The calculated band structure with SOC included is shown in Fig.~\ref{Ta3TeI7}(c). The material is a semiconductor with a bandgap about 0.48~eV. Although it is not a semimetal, by examining the band dispersion around the $k_z=\pi$ plane, one finds that there is indeed a nodal surface in that plane consisting of linear band-crossing points. This is more easily observed in the right panel of Fig.~\ref{Ta3TeI7}(c), which shows a zoom-in image for the dispersion along the generic path E-R. Along the path, the bands are non-degenerate (due to the broken $\mathcal{P}$), and a pair of bands must cross at R in the $k_z=\pi$ plane, hence a nodal surface is formed in that plane.

\section{Nodal surface in magnetic materials}\label{S4}
%As afore-emphasized, when the spin degrees of freedom are essential in material, the nonsymmorphic magnetic symmetry $\mathcal{T}S_{2z}$ can protect a nodal surface only if $\mathcal{PT}$ is violated. Except breaking $\mathcal{P}$, this requirement can also be satisfied with magnetic orders being developed perpendicular to the rotation axis of $S_{2z}$, and thus nodal surfaces can as well be protected by $\mathcal{T}S_{2z}$ in magnetic materials with such magnetic orders. Although both $\mathcal{T}$ and $S_{2z}$ are spontaneously broken, the combination $\mathcal{T}S_{2z}$ is preserved, since the spins aligning in the $xy$-plane are flipped by both $\mathcal{T}$ and $S_{2z}$, and therefore are intact by $\mathcal{T}S_{2z}$. Meanwhile, $\mathcal{PT}$ is definitely violated, because $\mathcal{P}$ preserves spin orientation, but $\mathcal{T}$ does not. On the other hand, once SOC is considered in crystals, the isotropy for magnetization is generically destroyed, and therefore it is natural to propose $S_{2z}$ symmetric materials with SOC, which can develop magnetic order in the $xy$-plane with our symmetry requirement being satisfied.

The discussion in the previous sections are about nonmagnetic materials, such that $\mathcal{T}$ is preserved. We have seen that $\mathcal{T}$ plays an important role in stabilizing the nodal surfaces. Then the question arises: \emph{Is it possible to realize protected nodal surfaces also in magnetic materials?}

Here we show that this is indeed possible, and the result also depends on whether the SOC in the system can be neglected or not.

First, consider the case in the absence of SOC. Without SOC, the spin and the orbital degrees of freedom are independent and can be considered as different subspaces. The spins can be oriented in any direction without affecting the orbital part of the wave-function. With a chosen spin polarization axis, the two spin channels are decoupled, and hence the bands for \emph{each} spin species can be \emph{effectively} regarded as for a spinless system. Therefore, for the states of one spin, all the crystalline symmetries are preserved~\cite{Wang2016,Chang2017}. Consequently, the analysis can be reduced to those in Sec.~\ref{S2}, which means that it is in principle possible to have class-I as well as class-II nodal surfaces in a magnetic system, if the symmetry requirements presented in Sec.~\ref{S2} are satisfied. Particularly, if the crystal structure possesses a two-fold screw axis $S_{2z}$, then there must exist a nodal surface (for each spin species) at $k_z=\pi$.

\begin{figure}[hbt]
\includegraphics[width=0.48\textwidth]{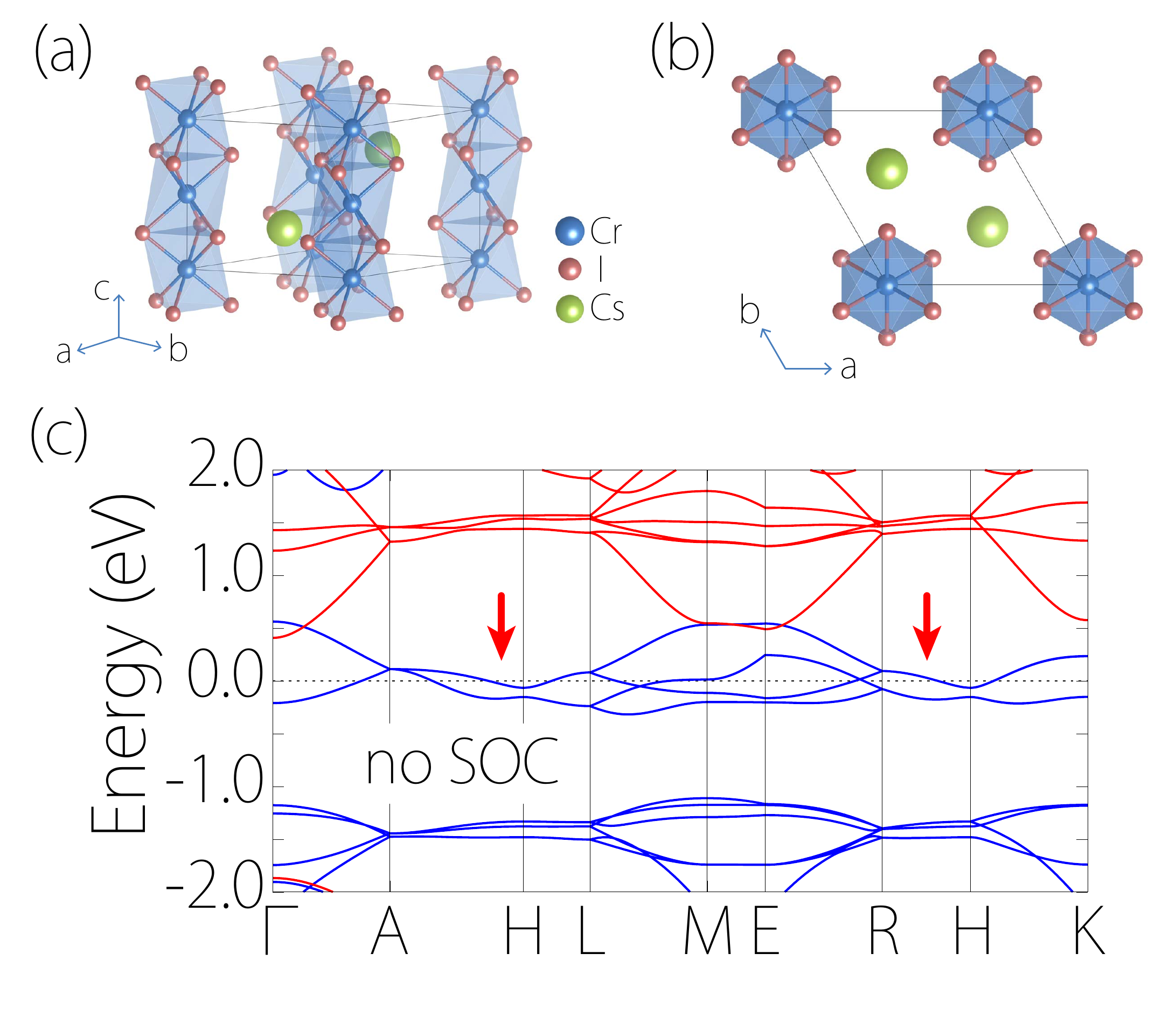}
\caption{(a) Side view and (b) top view of the crystal structure for CsCrI$_{3}$. (c) Calculated band structure for CsCrI$_{3}$ in the absence of SOC, showing a ferromagnetic state. The red and blue lines represent spin-up and spin-down bands, respectively. In the absence of SOC, the nodal surface is preserved, and the band structure does not depend on the direction of the magnetic moments. }
\label{CsCrI3}
\end{figure}

We demonstrate the above point using a concrete material example. We consider the material CsCrI$_{3}$, which is a member of the material family CsCr$X_{3}$ ($X=$ Cl, Br, I)~\cite{Guen1979Structure-ACSBSCaCC}. The material takes the BaVS$_{3}$-type structure, with space group $P6_{3}/mmc$ (No.~194), as shown in Fig.~\ref{CsCrI3}(a) and~\ref{CsCrI3}(b). The Cr atoms are surrounded by octahedra of $X$ atoms, forming one-dimensional chains along the $z$-axis, with octahedra sharing common faces. Those chains are arranged into a trigonal lattice in the $x$-$y$ plane, with the Cs atoms intercalated between the chains. Without considering the spin polarization, the lattice structure has a two-fold screw axis $S_{2z}$. The material assumes a ferromagnetic ground state, for which the magnetic moments are provided by the Cr atoms.
The band structure for CsCrI$_{3}$ without SOC is plotted in Fig.~\ref{CsCrI3}(c). One observes that the system is a half metal: The Fermi level crosses only the spin-up bands, whereas the spin-down bands are high in energy. Focusing on the spin-up bands that are near the Fermi level, one finds that there is indeed a nodal surface in the $k_z=\pi$ plane, as consistent with our above argument for the case without SOC. Due to the absence of SOC, orienting the magnetic moments along different directions will not affect the band structure, and hence will not affect the nodal surface. This is confirmed by our DFT results: We have repeated the calculation with different magnetic moment orientations (e.g., along the $x$-direction and the $z$-direction), and the obtained results are exactly the same as in Fig.~\ref{CsCrI3}(c).

Next, we consider the case in the presence of SOC. The spin and orbital degrees of freedom are tied with each other by SOC. It is generally not possible to label the bands as spin-up/spin-down. Hence, the symmetry operations need to act on both orbital and spin as a whole. Nevertheless,
we note that the arguments in Sec.~\ref{S3} rely on the existence of the composite symmetry $\mathcal{T}S_{2z}$, which can still be preserved although the individual $\mathcal{T}$ or $S_{2z}$ symmetry may be broken. If so, then the previous arguments in Sec.~\ref{S3} still apply and a nodal surface can be guaranteed. Explicitly, this means that: in the presence of SOC, a nodal surface in the $k_z=\pi$ plane can be protected by the combined symmetry $\mathcal{T}S_{2z}$ in conjunction with the absence of the $\mathcal{PT}$ symmetry in the system (note that $\mathcal{PT}$ may be preserved in certain antiferromagnet).

\begin{figure}[hbt]
\includegraphics[width=0.46\textwidth]{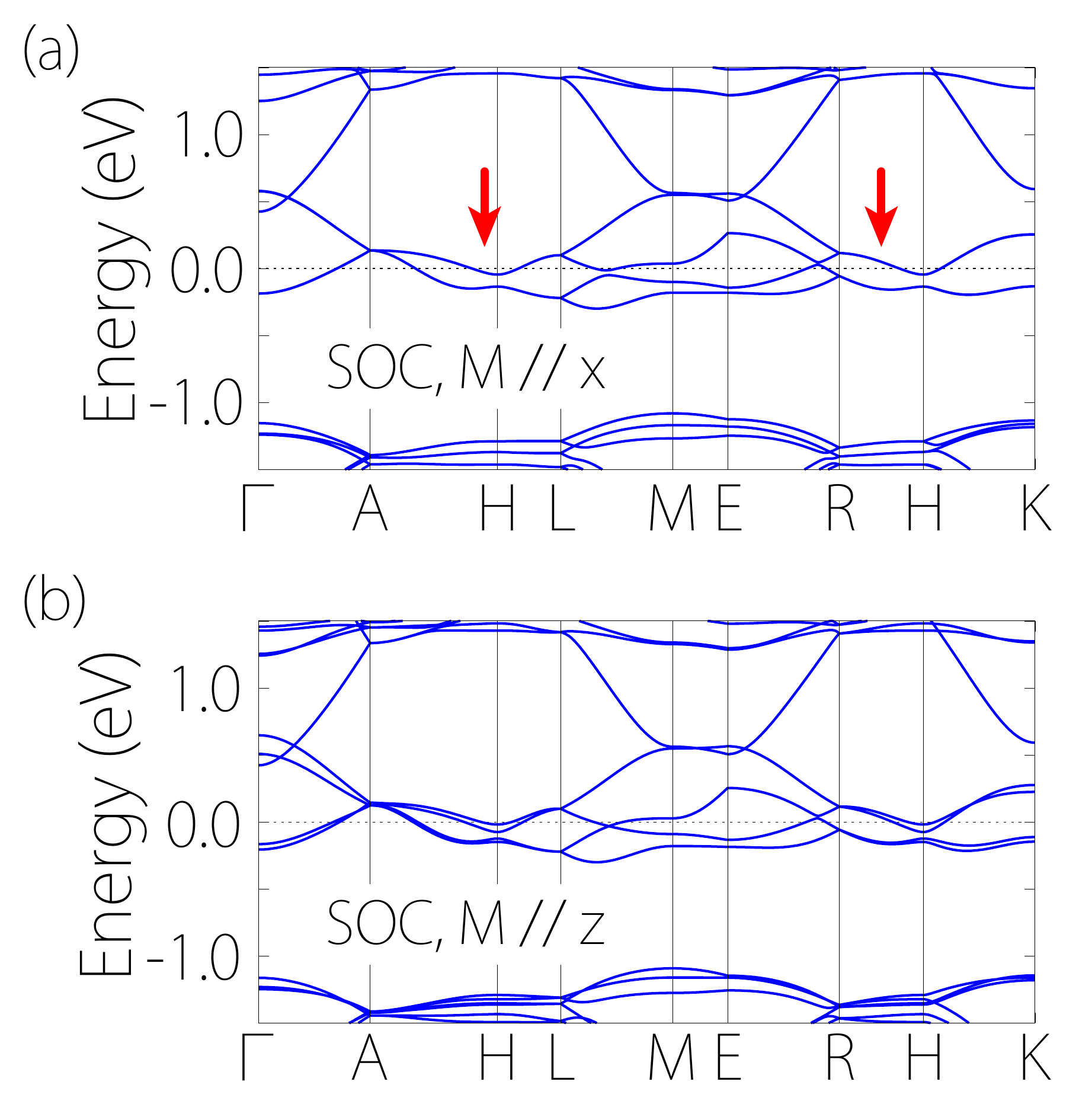}
\caption{Calculated band structures for CsCrI$_{3}$ in the presence of SOC, with magnetic moments along (a) the $\hat x$ direction and (b) the $\hat z$ direction. The nodal surface is preserved in (a) but not in (b).}
\label{CsCrI3_soc}
\end{figure}

We still take CsCrI$_{3}$ as an example and now include SOC in the DFT calculation. It is important to note that: (i) if the magnetic moments are aligned in a direction perpendicular to the $z$ direction, then the combined $\mathcal{T}S_{2z}$ symmetry is preserved (since the spins aligning in the $x$-$y$ plane are flipped by both $\mathcal{T}$ and $S_{2z}$, and therefore are intact by $\mathcal{T}S_{2z}$); and (ii) for other directions (e.g., along the $z$ direction), the $\mathcal{T}S_{2z}$ symmetry is broken. The $\mathcal{PT}$ symmetry is broken for both cases. Then according to our previous analysis, the nodal surface in the $k_z=\pi$ plane should be preserved in case (i) but not in case (ii). This is indeed confirmed by our DFT results. Figure~\ref{CsCrI3_soc}(a) shows that result with the moments aligned along the ${x}$-direction, where a nodal surface is observed in the $k_z=\pi$ plane. Figure~\ref{CsCrI3_soc}(b) shows that result when the moments are aligned along the ${z}$-direction, and one can see that the nodal surface is destroyed.

\section{Discussion and conclusion}\label{V}

We have discussed two classes of nodal surfaces in the absence of SOC. They are essentially different, namely the class-I surfaces are topologically charged, and two patches on the surfaces related by $\mathcal P$ have opposite charges to satisfy the Nielsen-Ninomiya no-go theorem~\cite{Nielsen1981,Nielsen1981a}; however, the class-II ones are mainly originated from the topological features of screw symmetry, and therefore each degenerate point belongs to a M\"{o}bius-strip band structure, all of which spread out as a torus (a $k_x$-$k_y$ sub-BZ).
The class-I surfaces are not essential, requiring band inversions in part of the BZ; whereas the class-II surfaces are essential and guaranteed to appear solely by the symmetry.
In addition, the shape and location of the class-I surfaces can vary in the BZ, and can be tuned, e.g. by lattice strain; whereas the the location of the class-II surface is fixed at the BZ boundary plane as a result of the combination of time reversal and screw rotation. The nodal surfaces in the examples discussed in Sec.~\ref{S3} and \ref{S4} are generalizations of the class-II surfaces, so they share similar characteristic features of the class-II surfaces listed above.

We mention that both classes do not lead to any special boundary gapless modes. Although the first class has the nontrivial $\mathbb{Z}_2$ topological charge, similar to Weyl semimetals, the codimensionality of nodal surfaces or the dimensionality of sub topological insulators is zero, impossible to generate any boundary state. It may be noteworthy that each nodal surface considered here is a torus in the BZ, which cannot be continuously deformed to be a point, and hence should be distinguished from a nodal surface that is topologically a sphere deformable to a single point in the BZ~\cite{Turker2017}. (Note that for class-I surfaces, a pair of surfaces may merge into a single sphere when the system is strongly deformed.) We also mention that although the 0D $\mathbb{Z}_2$ charge does not produce any topological surface state, it is possible that additional 1D or 2D topological charges (having codimensions 1 and 2) may exist and lead to nontrivial surface states in these systems.

As we have mentioned, the nodal surfaces are distinct from the usual Fermi surfaces because it is formed by the crossing of two bands. As a result, the low-energy electrons acquire an intrinsic pseudo-spin degree of freedom, different from the usual Schr\"{o}dinger fermions. Near a relatively flat nodal surface, the low-energy effective model may be written as
\begin{equation}\label{Heff}
\mathcal{H}_\text{eff}=v_z q_z\sigma_z,
\end{equation}
where $v_z$ is the Fermi velocity, $q_z$ measures the deviation from the nodal surface along the surface normal direction, and the Pauli matrix $\sigma_z$ represents the pseudo-spin, corresponding to the two crossing bands. Equation~(\ref{Heff}) effectively describes 1D massless Dirac fermions, where the pseudospin is locked to the momentum. For transport along the $z$-direction, we may expect interesting properties such as the enhanced mobility due to the suppression of back-scattering (for states around a single nodal surface) from scatterers that preserve the pseudospin.

\begin{figure}[hbt]
	\includegraphics[width=0.47\textwidth]{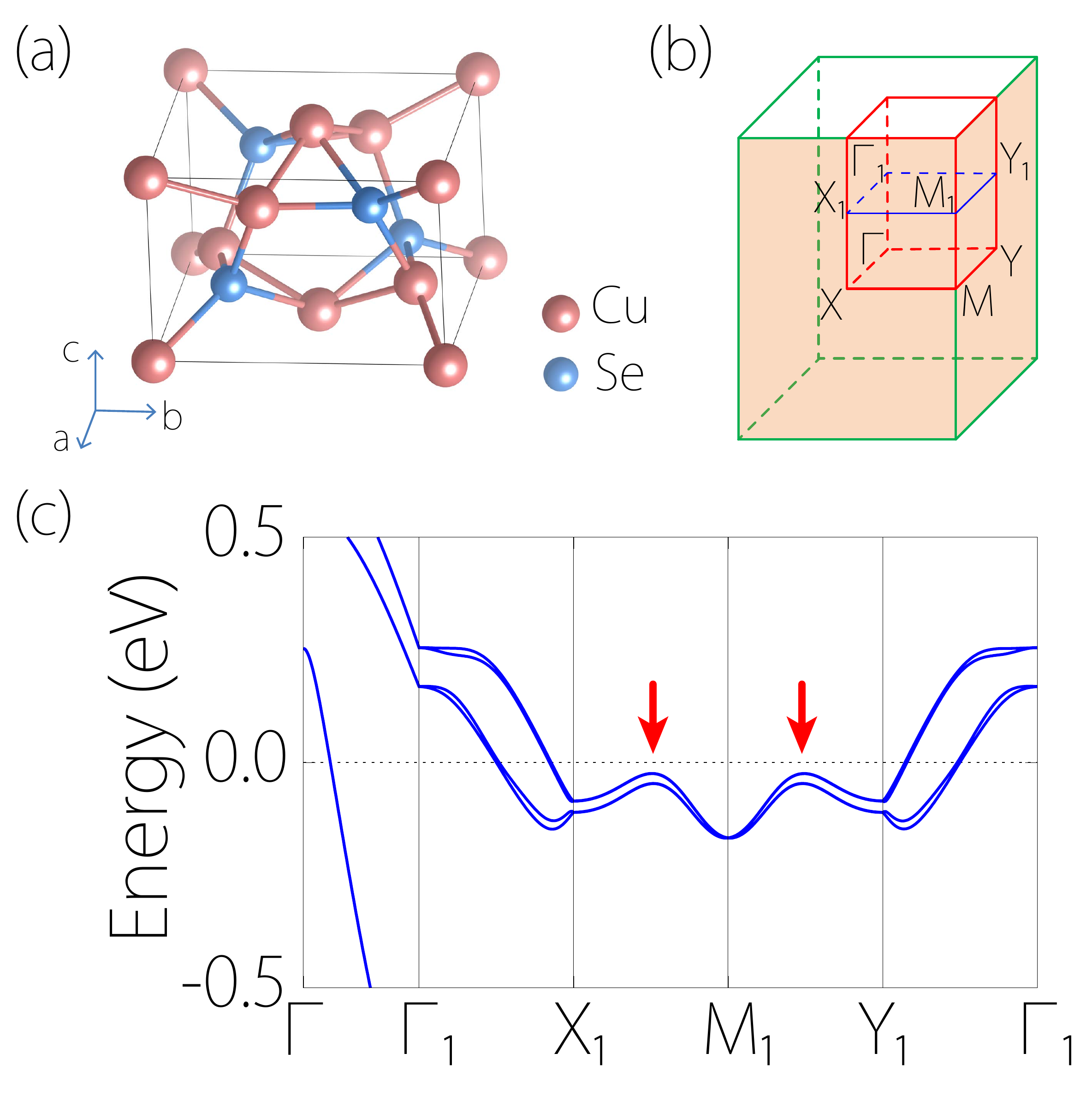}
	\caption{(a) Crystal structure of Cu$_{3}$Se$_{2}$ and (b) the corresponding Brillouin zone. In (b), $\Gamma_{1}$, X$_{1}$, Y$_{1}$, and M$_{1}$ label the four $k$-points that are above the points $\Gamma$, X, Y, and M, and are located in the $k_{z}=\pi/2c$ plane. (c) Band structure of Cu$_{3}$Se$_{2}$ (with SOC) along some generic paths. The red arrows indicate the band degeneracy along the paths on the two orthogonal nodal surfaces.}
	\label{Cu3Se2}
\end{figure}

Regarding the class-II surfaces, if there are multiple two-fold screw axis, then it is possible to have multiple nodal surfaces on the BZ boundary planes. One example is shown in Fig.~\ref{Cu3Se2} for the material Cu$_{3}$Se$_{2}$ known as the mineral umangite, with the space group of $P\overline{4}2_{1}m$ (No.~113) \cite{Morimoto1966Crystal-S,Chakrabarti1981Cu-JoPE}. As a typical intermetallic compounds, Cu atoms are bonded closely both with one another and with Se atoms. Two kinds of Cu atoms exist in the structure, with Cu(I) at the $2a$ position $(0,0,0)$ and Cu(II) at the $4e$ position $(x, \frac{1}{2}-x, z)$ with $x = 0.145$ and $z = 0.763$. The material experiences a transition to an antiferromagnetic state at about 50 K \cite{Okamoto1969Electrical-JJoAP}. Here, we focus on its paramagnetic phase above 50 K. The band structure of Cu$_{3}$Se$_{2}$ is plotted in Fig.~\ref{Cu3Se2}(c). The space group $P\overline{4}2_{1}m$ contains two perpendicular screw axis $S_{2x}=\{C_{2x}|\frac{1}{2}\frac{1}{2}0\}$ and $S_{2y}=\{C_{2y}|\frac{1}{2}\frac{1}{2}0\}$. With preserved $\mathcal{T}$, two mutually orthogonal nodal surfaces can be found in the $k_{x}=\pi$ and $k_{y}=\pi$ planes [as illustrated in Fig.~\ref{Cu3Se2}(b)]. This is indeed observed in the band structure in Fig.~\ref{Cu3Se2}(c), which shows that the non-degenerate bands along the generic paths $\Gamma_{1}$-X$_{1}$ and $\Gamma_{1}$-Y$_{1}$ cross at the two nodal surfaces. A material example with three nodal surfaces is also presented in Appendix C.

Finally, we mention that as a good NSSM material, the nodal surface should be close to the Fermi level and formed by the crossing between conduction and valence bands. This condition imposes constraints regarding the electron filling of the bands. The recent work by Watanabe \emph{et al.}~\cite{Watanabe2016} has studied the detailed filling constraints for realizing semimetal states for nonsymmorphic space groups, which will offer useful guidelines for searching NSSMs. In this work, we have identified several good candidate materials that satisfy the condition. However, for some cases, like the case in the presence of SOC, the examples we show are not ideal. For example, Ta$_3$TeI$_7$ discussed in Sec.~\ref{S3} is actually a semiconductor. Nevertheless, it serves the purpose to illustrate that a nodal surface can indeed appear when the proposed symmetry requirements are satisfied. Based on our theory, we expect better NSSM materials to be discovered in future studies.

In conclusion, we have theoretically investigated the NSSMs which host robust nodal surfaces formed by the crossing of two bands close to the Fermi level. We clarify the symmetry/topology-protection of the nodal surfaces. In the absence of SOC, we identify two classes of nodal surfaces which are protected by different mechanisms. The class-I surfaces are protected by a combination of inversion, sublattice, and time reversal symmetries and are characterized by a $\mathbb{Z}_2$ index. The class-II surfaces are protected by a two-fold screw axis and the time reversal symmetry. The two classes differ in several aspects including the surface shape, location, and robustness against perturbation. The inclusion of SOC generally gaps the class-I surface, and we show that the class-II surface may be preserved provided that the inversion symmetry is broken. Furthermore, we generalize the analysis to magnetic materials. We find that class-II surfaces can exist in magnetic materials both without and with SOC, given that certain magnetic group symmetry requirements are satisfied. We have identified several concrete NSSM material examples, which will facilitate the experimental exploration of the intriguing properties of NSSMs.

\begin{acknowledgements}
The authors thank T. Bzdu\v{s}ek, X. C. Wu, and D. L. Deng for valuable discussions and comments on the manuscript. This work was supported by the Singapore Ministry of Education Academic Research Fund Tier 1 (SUTD-T1-2015004) and Tier 2 (MOE2015-T2-2-144). We acknowledge computational support from the Texas Advanced Computing Center and the National Supercomputing Centre Singapore.
\end{acknowledgements}

\ \
\par
\ \
\clearpage
\begin{appendix}
\renewcommand{\theequation}{A\arabic{equation}}
\setcounter{equation}{0}
\renewcommand{\thefigure}{A\arabic{figure}}
\setcounter{figure}{0}
\renewcommand{\thetable}{A\arabic{table}}
\setcounter{table}{0}

\begin{widetext}
\section{Tight-binding model and $\mathbb{Z}_2$ invariant for QGN(1,2)}\label{TB-model}

As discussed in the main text, the low-energy physics mainly comes from electrons hopping among the eight $sp^2$ sites, and therefore the Hamiltonian of the tight-binding model, $H=\int d^3 k~\psi^\dagger_{\bm k}\mathcal{H}(\bm{k})\psi_{\bm k}$ with $\psi_{\bm k}=(c_1,c_2,\cdots,c_8)^T$, describes an eight-band theory. After a gauge transformation to make the Hamiltonian periodic in $k_x$ and $k_y$, we have

	\begin{equation}\label{TB}
	\small{
		\mathcal{H}(\bm k)=\begin{pmatrix}
		0    & 2t_1 \cos(\frac{1}{2}k_z) &   0   & 0 & 0 & 0 & 0 & t_2 e^{ik_x} \\
		2t_1 \cos(\frac{1}{2}k_z) & 0 & t_2 e^{ik_y} & 0 & 0 & 0 & 0 & 0\\
		0   & t_2 e^{-ik_y} & 0 & 2t_1\cos(\frac{1}{2}k_z) & 0 & 0 & 0 & 0\\
		0 & 0 & 2t_1\cos(\frac{1}{2}k_z) & 0 & t_2 e^{ik_x} & 0 & 0 & 0\\
		0 & 0 & 0 & t_2 e^{-ik_x} & 0 & 2t_1 \cos(\frac{1}{2}k_z) & 0 & 0\\
		0 & 0 & 0 & 0 & 2t_1 \cos(\frac{1}{2}k_z) & 0 & t_2 e^{-ik_y} & 0 \\
		0 & 0 & 0 & 0 & 0 & t_2 e^{ik_y} & 0 &  2t_1 \cos(\frac{1}{2}k_z)\\
		t_2 e^{-ik_x} & 0 & 0 & 0 & 0 & 0 &  2t_1 \cos(\frac{1}{2}k_z) & 0
		\end{pmatrix}}.
	\end{equation}
\end{widetext}
Here the wave-vectors are measured in the respective inverse lattice constants, and $t_1$ and $t_2$ are the real hopping amplitudes from site 1 to site 2 and from site 2 to site 3, respectively [see Fig.~\ref{QGN}(b)].

Performing the combined unitary transformation $U_{PT}U_{\Gamma}$ with
\begin{equation}
U_{\Gamma}=\renewcommand{\arraystretch}{0.6}\begin{pmatrix}
1 & 0 & 0 & 0 & 0 & 0 & 0 & 0\\
0 & 0 & 0 & 0 & 0 & 0 & 1 & 0\\
0 & 0 & 1 & 0 & 0 & 0 & 0 & 0\\
0 & 0 & 0 & 0 & 1 & 0 & 0 & 0\\
0 & 0 & 0 & 1 & 0 & 0 & 0 & 0\\
0 & 0 & 0 & 0 & 0 & 1 & 0 & 0\\
0 & 1 & 0 & 0 & 0 & 0 & 0 & 0\\
0 & 0 & 0 & 0 & 0 & 0 & 0 & 1
\end{pmatrix}
\end{equation}
and
\begin{equation}
U_{PT}=\frac{1}{\sqrt{2}}\renewcommand{\arraystretch}{0.6}\begin{pmatrix}
1 & 0 & 0 & i & 0 & 0 & 0 & 0\\
0 & 1 & i & 0 & 0 & 0 & 0 & 0\\
0 & i & 1 & 0 & 0 & 0 & 0 & 0\\
i & 0 & 0 & 1 & 0 & 0 & 0 & 0\\
0 & 0 & 0 & 0 & 1 & 0 & 0 & i\\
0 & 0 & 0 & 0 & 0 & 1 & i & 0\\
0 & 0 & 0 & 0 & 0 & i & 1 & 0\\
0 & 0 & 0 & 0 & i & 0 & 0 & 1
\end{pmatrix},
\end{equation}
the Hamiltonian becomes anti-diagonal [i.e., $\mathcal{H}(\bm{k})=\mathrm{antidiag}(Q(\bm{k}),Q^\dagger(\bm{k}))$] with the upper right lock being
\begin{equation}
%\small{
Q(\bm{k})=\begin{pmatrix}
t_2\sin k_x & 0 & 2t_1\cos\frac{k_z}{2} & t_2\cos k_x\\
0 & t_2\cos k_y & t_2\sin k_y & 2t_1\cos\frac{k_z}{2}\\
2t_1\cos\frac{k_z}{2} & -t_2\sin k_y & t_2\cos k_y & 0\\
t_2\cos k_x & 2t_1\cos\frac{k_z}{2} & 0 & -t_2\sin k_x
\end{pmatrix}.%}
\end{equation}

\begin{figure}[hbt]
	\includegraphics[width=0.42\textwidth]{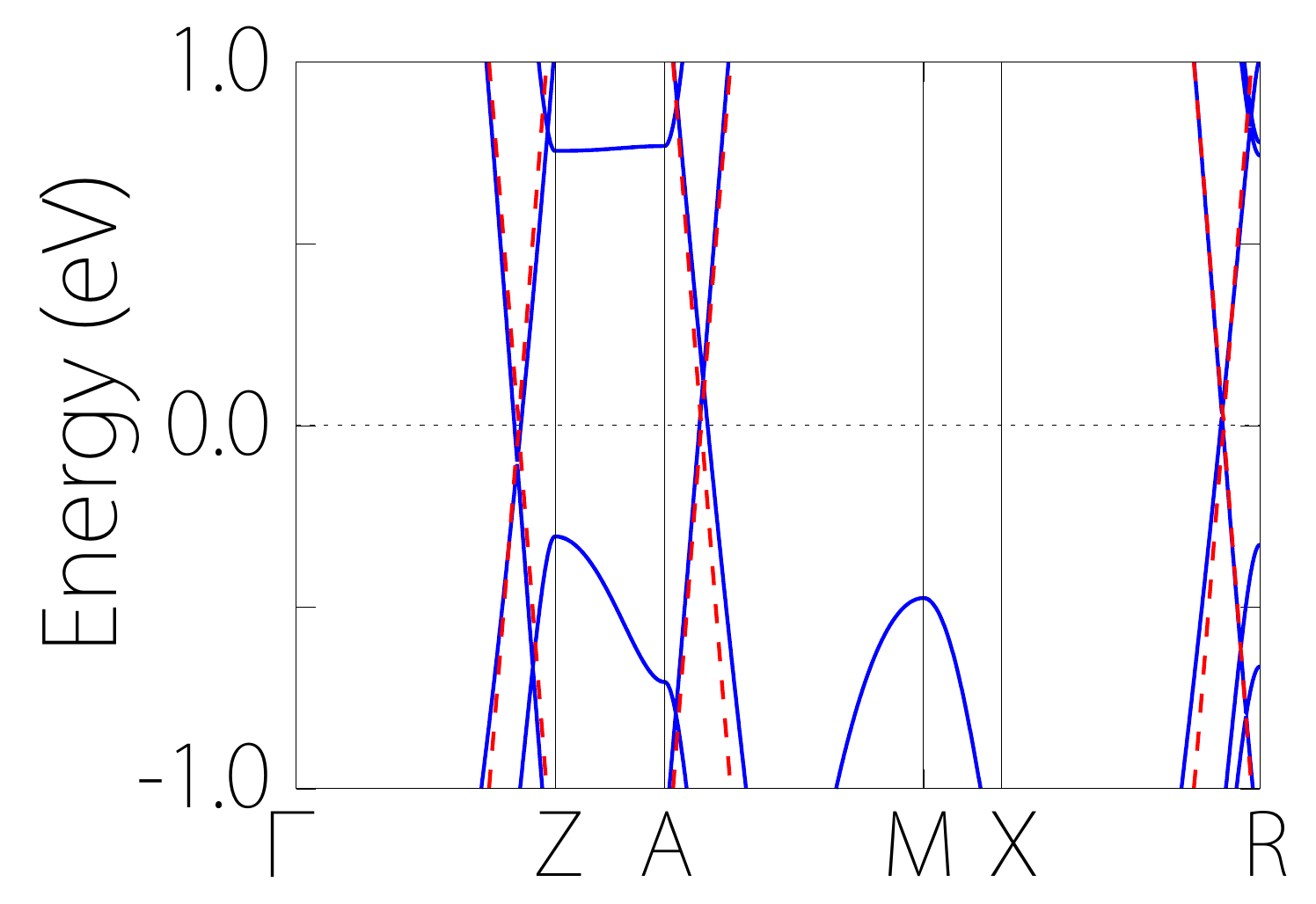}
	\caption{Comparison of band structures of the eight-band tight binding model in Eq.~\ref{TB} (red dashed lines) and the DFT result (blue solid lines). In the tight-binding model, the values $t_1=2.95$ eV and $t_2=1.30$ eV are obtained from fitting the low-energy DFT band structure. }
	\label{QGN12_TB}
\end{figure}

Then it is found that $\mathrm{Det}Q(\bm{k})=4t_1^4(\cos k_z+1)^2-t_2^4$. If $|t_2/2t_1|<1$, then there are two nodal surfaces with $k_z=\pm \arccos (t_2^2/2t_1^2-1)$ related by $\mathcal{T}$ or $\mathcal{P}$ symmetry. It is easy to check that $\mathrm{Det}Q(k_z=0)>0$ because $|t_2/2t_1|<1$ ($t_1=2.95$ eV and $t_2=1.30$ eV, as obtained from the fitting of DFT result in Fig.~\ref{QGN12_TB}); whereas $\mathrm{Det}Q(k_z=\pi)<0$. Thus, both nodal surfaces have the nontrivial topological charge according to the formula in Eq.~(\ref{Z2}).

\section{First-principles calculation method}
The first-principle calculations are based on the DFT, as implemented in the Vienna \emph{ab initio} simulation package~\cite{kresse1993VASP, kresse1996VASP}.  The projector augmented wave method was adopted~\cite{blochl1994PAW}. The generalized gradient approximation (GGA) with the Perdew-Burke-Ernzerhof (PBE) realization~\cite{PBE1996PBE} was adopted for the exchange-correlation potential. For all calculations, the energy and force convergence criteria were set to be $10^{-5}$~eV and $10^{-2}$~eV/\text{\AA}, respectively. The BZ sampling was performed by using $k$ grids with a spacing of $2\pi \times 0.02$ \AA$^{-1}$ within a $\Gamma$-centered sampling scheme. As the transition metal \emph{d} orbitals may have notable correlation effects, we have validated our DFT results by the GGA+U method~\cite{Dudarev1998} (see Appendix~\ref{plusU}). The key features are found to be qualitatively the same as the GGA results. Hence in the main text, we focus on the GGA results. For K$_{6}$YO$_{4}$, the optimized lattice parameters ($a = 9.59~\text{\AA}$, $c = 6.69~\text{\AA}$) were used for the band structure calculation. For the other materials discussed here, the experimental values of their respective lattice parameters were used in the calculation.

\section{SOC effect on class-I surfaces}\label{appC}
\begin{figure}[!hbt]
\includegraphics[width=0.48\textwidth]{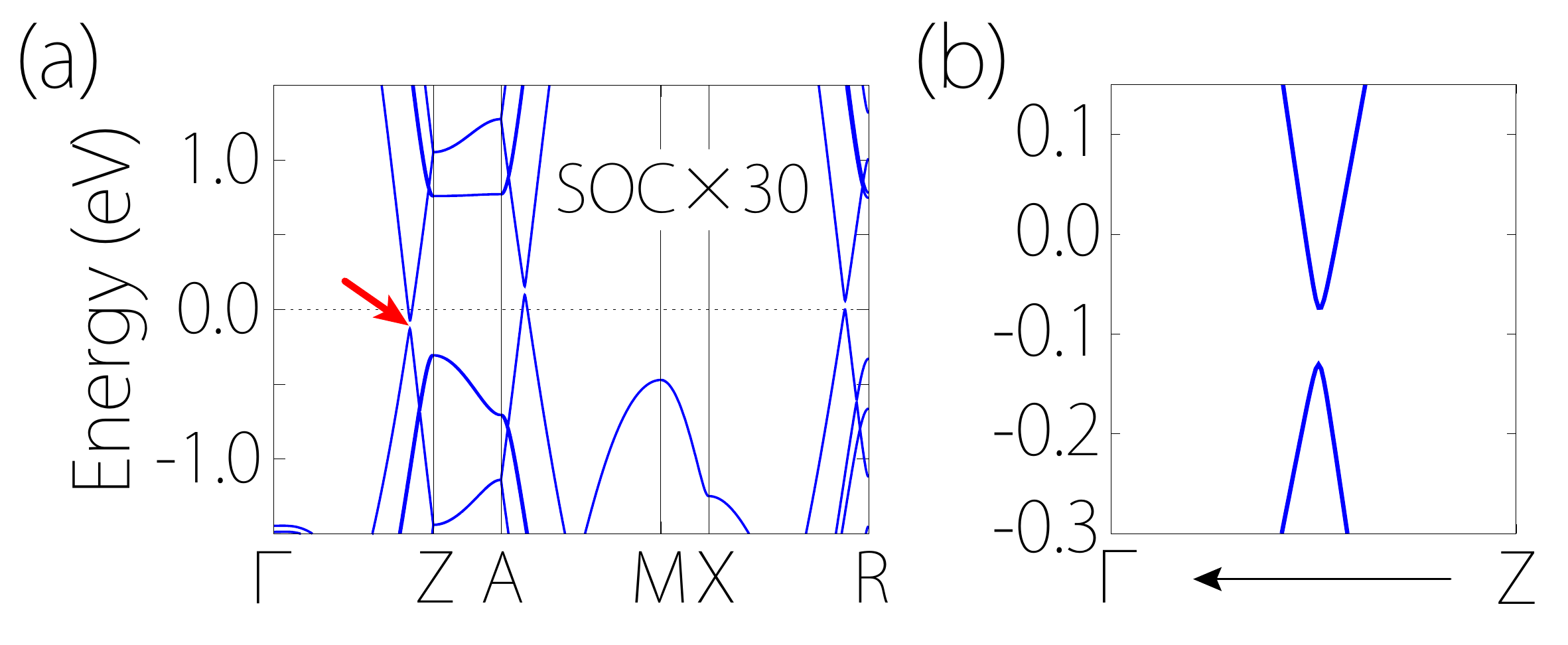}
\caption{(a) Band structure for QGN(1,2) with SOC. The SOC strength is artificially enhanced by 30 times, in order to show that the SOC gaps the original nodal surfaces. (b) Enlarged view around the original band crossing, corresponding to the region as indicated by the red arrow in (a).}
\label{QGN12_soc}
\end{figure}

Here we consider the SOC effect on the class-I nodal surfaces. Take the QGN(1,2) discussed in Sec.~\ref{2A} as an example. Since the carbon is a light element with very weak SOC. The band structure with SOC shows negligible difference from that in Fig.~\ref{QGN}(d) without SOC. In order to more clearly demonstrate that SOC in fact gaps the nodal surface, we artificially enhance the SOC strength in the DFT calculation by 30 times. The obtained band structure is plotted in Fig.~\ref{QGN12_soc}. One clearly observes that a gap is opened by SOC at the original band-crossing point, destroying the class-I nodal surface.

\section{Material example with three nodal surfaces}

\begin{figure}[!hbt]
	\includegraphics[width=0.46\textwidth]{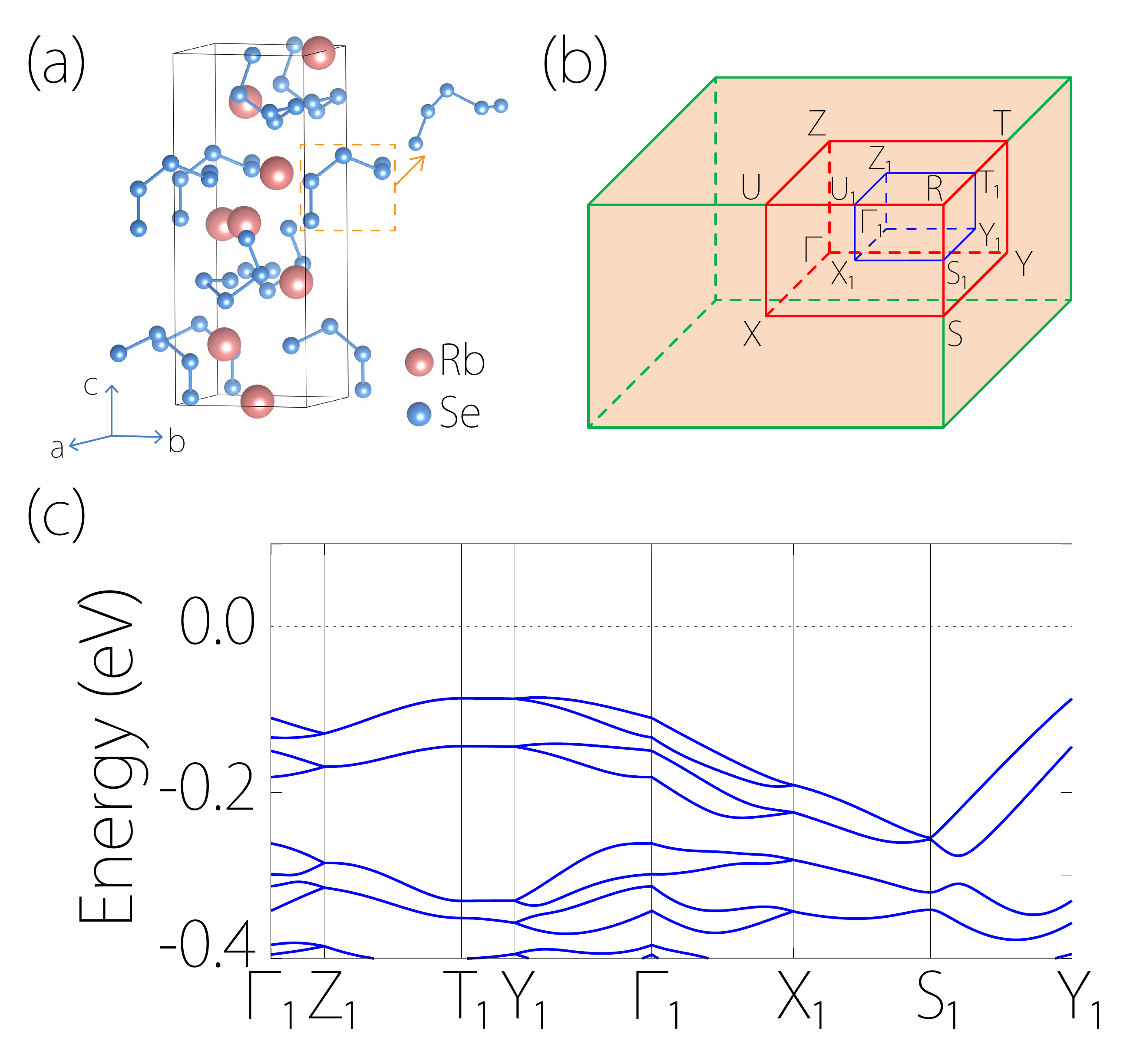}
	\caption{(a) Crystal structure of Rb$_{2}$Se$_{5}$ and (b) the corresponding Brillouin zone. The insert in (a) shows the helical Se$_{5}^{2-}$ chain. In (b), $\Gamma_{1}$ is the body center of 1/8 Brillouin zone; X$_{1}$, Y$_{1}$, and Z$_{1}$ are the face centers of the 1/8 Brillouin zone; U$_{1}$, T$_{1}$, and S$_{1}$ are the mid-points of the paths U-R, T-R, and S-R, respectively. (c) Band structure of Rb$_{2}$Se$_{5}$ with SOC.}
	\label{Rb2Se5}
\end{figure}

In Sec.~\ref{V}, we have shown that the compound Cu$_{3}$Se$_{2}$ with two perpendicular screw axis can host two orthogonal nodal surfaces on the BZ boundary planes. Here we consider another example material Rb$_{2}$Se$_{5}$ with three perpendicular nodal surfaces on the BZ boundary planes \cite{Boettcher1979Synthesis-ZfK1}. Its crystal structure has the space group $P2_{1}2_{1}2_{1}$ (No.~19), which has three two-fold screw axes perpendicular to each other: $S_{2x}=\{C_{2x}|\frac{1}{2}\frac{1}{2}0\}$, $S_{2y}=\{C_{2y}|0\frac{1}{2}\frac{1}{2}\}$, and $S_{2z}=\{C_{2z}|\frac{1}{2}0\frac{1}{2}\}$. The time reversal symmetry is preserved due to the absence of magnetic ordering. There are four helical Se$^{2-}_5$ chains in the unit cell, and the Rb atoms are surrounded by the Se chains and related by the screw axis.

The calculated band structure for Rb$_{2}$Se$_{5}$ with SOC included is shown in Fig.~\ref{Rb2Se5}(c) (only some generic paths are shown here). We find that the material is a semiconductor with a bandgap about 0.75~eV. Nevertheless, band crossings appear in both valence and conduction bands. By examining the band dispersion on the BZ boundary planes, one can find that the bands cross at these planes in pairs. For example, the non-degenerate bands along the three generic paths $\Gamma_{1}$-X$_{1}$, $\Gamma_{1}$-Y$_{1}$, $\Gamma_{1}$-Z$_{1}$ all cross at the BZ boundary planes. Hence there are three mutually orthogonal nodal surfaces as illustrated in Fig.~\ref{Rb2Se5}(b).

\section{Hubbard U correction}\label{plusU}

\begin{figure*}[!hbt]
	\includegraphics[width=1\textwidth]{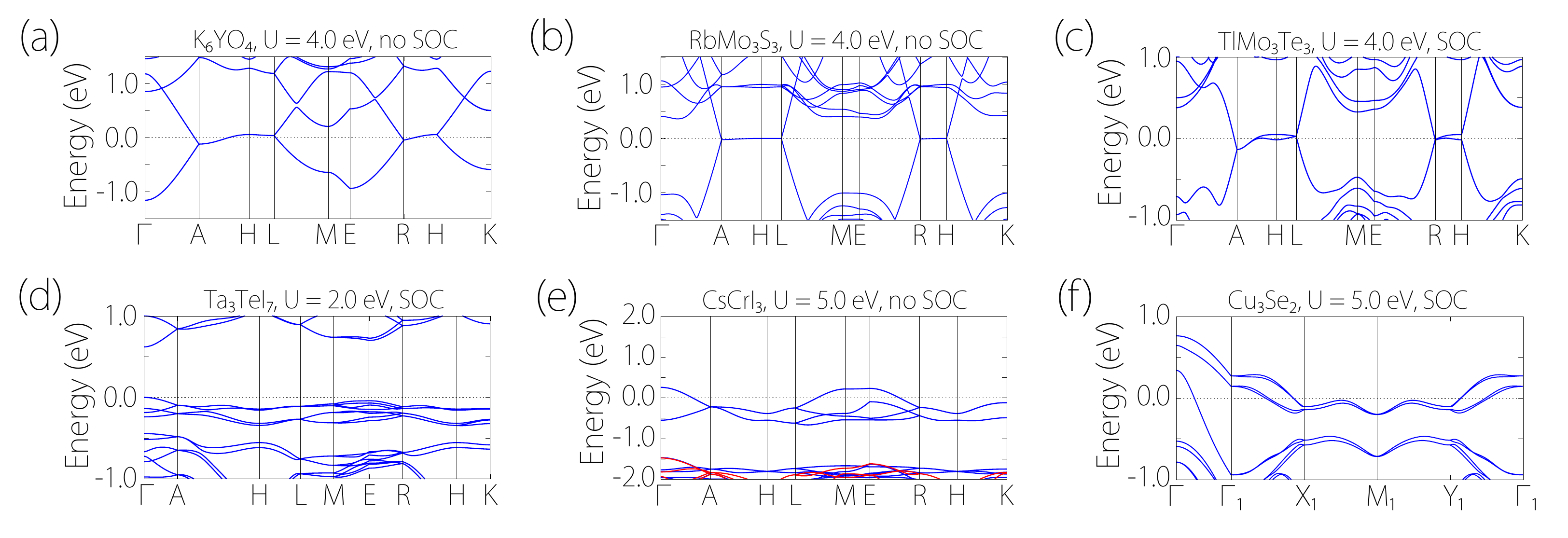}
	\caption{Effects of the Hubbard U correction on band structures of (a) K$_{6}$YO$_{4}$ (without SOC) with U=4.0~eV, (b) RbMo$_{3}$S$_{3}$ (without SOC) with U=4.0~eV, (c) TlMo$_{3}$Te$_{3}$ (SOC included) with U=4.0~eV, (d) Ta$_{3}$TeI$_{7}$ (SOC included) with U=2.0~eV, (e) CsCrI$_{3}$ (without SOC) with U=5.0~eV, and (f) Cu$_{3}$Se$_{2}$ (SOC included) with U=5.0~eV.}
	\label{ldau}
\end{figure*}

To test the correlation effects of the transition metal \emph{d}-orbitals, we performed the GGA+U calculations, by taking into account the on-site Coulomb interaction. The U values proper for each transition metal element have been tested. Figure~\ref{ldau} shows the representative results for the band structures of K$_{6}$YO$_{4}$, RbMo$_{3}$S$_{3}$, TlMo$_{3}$Te$_{3}$, Ta$_{3}$TeI$_{7}$, CsCrI$_{3}$, and Cu$_{3}$Se$_{2}$. One can observe that the results for all these compounds show little change in comparison with the GGA results. Hence, we focus on the GGA results in the main text.

\end{appendix}

%\bibliography{surface}

\end{document}